\documentclass[letterpaper, 10 pt, conference]{ieeeconf}%
\usepackage{graphics}
\usepackage{epsfig}
\usepackage{mathptmx}
\usepackage{slashbox}
\usepackage{times}
\usepackage{amsmath}
\usepackage{amssymb}
\usepackage{amsfonts}
\usepackage{graphicx}
\usepackage{eurosym}
\usepackage[usenames, dvipsnames]{color}%
\providecommand{\U}[1]{\protect\rule{.1in}{.1in}}
\IEEEoverridecommandlockouts
\renewcommand{\baselinestretch}{1}
\begin{document}

\title{{\LARGE \textbf{Time and Energy-Optimal Lane Change Maneuvers for Cooperating
Connected and Automated Vehicles$^{\star}$}}}
\author{Rui Chen, Christos G. Cassandras and Amin Tahmasbi-Sarvestani \thanks{
$^{\star}$Supported by Honda R\&D Americas. Also supported in part by NSF
under grants ECCS-1509084, CNS-1645681, and DMS-1664644, by AFOSR under grant
FA9550-15-1-0471, by ARPAE's NEXTCAR program under grant DE-AR0000796, and by
the MathWorks.} \thanks{The first two authors are with the Division of Systems
Engineering and Center for Information and Systems Engineering, Boston
University, Brookline, MA 02446 \texttt{{\small \{ruic,cgc\}@bu.edu}}. The
third author was with Honda R\&D Americas, Inc., 2420 Oak Valley Drive, Ann
Arbor, MI 48103 \texttt{{\small \{atahmasbi\}@hra.com}}.} }
\maketitle

\begin{abstract}
We derive optimal control policies for a Connected and Automated Vehicle (CAV)
cooperating with neighboring CAVs to implement a highway lane change maneuver.
We optimize the maneuver time and subsequently minimize the associated energy
consumption of all cooperating vehicles in this maneuver. We prove structural
properties of the optimal policies which simplify the solution derivations and
lead to analytical optimal control expressions. The solutions, when they
exist, are guaranteed to satisfy safety constraints
for all vehicles involved in the maneuver. Simulation results show the
effectiveness of the proposed solution and significant performance
improvements compared to maneuvers performed by human-driven vehicles.

\end{abstract}

\section{Introduction}

Advances in next generation transportation system technologies and the
emergence of Connected and Automated Vehicles (CAVs), also known as
\textquotedblleft autonomous vehicles\textquotedblright, have the potential to
drastically improve a transportation network's performance in terms of safety,
comfort, congestion reduction and energy efficiency. In highway driving, an
overview of automated intelligent vehicle-highway systems was provided in
\cite{varaiya1993smart} with more recent developments mostly focusing on
autonomous car-following control \cite{zhao2018accelerated}%
,\cite{wang2016cooperative},\cite{wang2015game}. Automating a lane change
maneuver remains a challenging problem which has attracted increasing
attention in recent years \cite{nilsson2015longitudinal},\cite{bax2014road}%
,\cite{you2015trajectory},\cite{werling2010optimal}.

The basic architecture of an automated lane-change maneuver can be divided
into the strategy level and the control level \cite{bevly2016lane}. The
strategy level generates a feasible (possibly optimal in some sense)
trajectory for a lane-change maneuver. The control level is responsible for
determining how vehicles track the aforementioned trajectory. For example,
\cite{you2015trajectory} adopts such an architecture for an automated
lane-change maneuver, but does not provide an analytical solution and assumes
that there are no other vehicles in the left lane (the lane in which the
controllable vehicle ends up after completing the maneuver). In
\cite{nilsson2017lane}, background vehicles are included in the left lane and
the goal is to check whether there exists a lane-change trajectory or not; if
one exists, the controllable vehicle will then track this trajectory. A
similar approach is taken in \cite{luo2016dynamic} with the trajectory being
updated during the maneuver based on the latest surrounding information. In
these papers, only one vehicle can be controlled during the maneuver and no
analytical solutions are provided.

The emergence of CAVs brings up the opportunity for cooperation among vehicles
traveling in both left and right lanes in carrying out an automated
lane-change maneuver \cite{bevly2016lane},\cite{mahjoub2017learning},\cite{kazemi2018learning}. Such cooperation presents several
advantages relative to the two-level architecture mentioned above. In
particular, when controlling a single vehicle and checking on the feasibility
of a maneuver depending on the state of the surrounding traffic, as in
\cite{kamal2013model},\cite{katriniok2013optimal}, the maneuver may be
infeasible without the cooperation of other vehicles, especially under heavier
traffic conditions. In contrast, a cooperative architecture can allow multiple
interacting vehicles to implement controllers enabling a larger set of
maneuvers. This cooperative behavior can also improve the throughput,
hence reducing the chance of congestion. Feasible, but not necessarily
optimal, vehicle trajectories for cooperative multi-agent lane-changing
maneuvers are derived in \cite{lam2013cooperative}. The case of multiple
cooperating vehicles simultaneously changing lanes is considered in
\cite{li2017optimal} with the requirement that all vehicles are controllable
and their velocities prior to the lane change are all the same. First,
vehicles with a lower priority must adjust their positions in their current
lane and give way to those with a higher priority so as to avoid collisions.
Then, a lane changing optimal control problem is solved for each vehicle
without considering the usual safe distance constraints between vehicles. This
\textquotedblleft progressively constrained dynamic
optimization\textquotedblright\ method facilitates a numerical solution to the
underlying optimal control problem at the expense of some loss in performance.

\begin{figure}[pt]
\centering
\includegraphics[angle=-90,origin=c,scale=0.25]{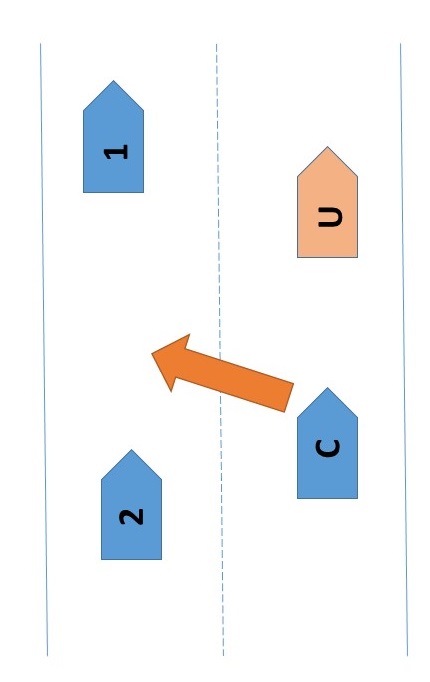}
\vspace*{-\baselineskip} \vspace*{-\baselineskip}\caption{The basic lane
changing maneuver process.}%
\label{maneuver_process}%
\end{figure}

Our goal is to provide an optimal solution for the maneuver in Fig.
\ref{maneuver_process}, in which the controlled vehicle $C$ attempts to
overtake an uncontrollable vehicle $U$ by using the left lane to pass. In this
case, the initial velocities of all vehicles can be different and arbitrary.
The overall lane changing and passing maneuver consists of three
steps: $(i)$ The target vehicle $C$ moves to the left lane, $(ii)$ $C$ moves
faster than $U$ (and possibly other vehicles ahead of it) while on the left
lane, $(iii)$ $C$ moves back to the right lane. The first step is further
subdivided into two parts. First, vehicle $C$ adjusts its position in the
current lane to prepare for a lane shift, while vehicles $1$ and $2$ in Fig.
\ref{maneuver_process} cooperate to create space for $C$ in the left lane.
Next, the latitudinal lane shift of $C$ takes place. In this paper, we limit
ourselves to the first part of step $(i)$. Our objective is to minimize both
the maneuver time and the energy consumption of vehicles $C$, $1$ and $2$
which are all assumed to share their state information. We also impose a hard
safe distance constraint between all adjacent vehicles located in the same
lane, as well as constraints due to speed and acceleration limits imposed on
all vehicles. We first determine a minimum feasible time for the maneuver (if
one exists) and associated terminal positions for vehicles $C$, $1$ and $2$.
We then solve a fixed terminal time decentralized optimal control problem for
each of the three vehicles. We derive several properties of the optimal
solution which facilitate obtaining explicit analytical solutions, hence
leading to real-time implementability. Our approach applies to a wider range
of scenarios relative to those in \cite{nilsson2017lane},\cite{luo2016dynamic}%
,\cite{kamal2013model},\cite{katriniok2013optimal} and incorporates the safety
distance constraint not included in \cite{lam2013cooperative} and
\cite{li2017optimal}.

The rest of this paper is organized as follows. Section II formulates the
lane-change maneuver problem. In Section III, a complete optimal control
solution is obtained. Section IV provides simulation results for several
representative examples and we conclude with Section V.

\section{Problem Formulation}

We define $x_{i}(t)$ to be the longitudinal position of vehicle $i$ along its current lane
measured with respect to a given origin, where we use $i=1,2,C,U$. Similarly,
$v_{i}(t)$ and $u_{i}(t)$ are vehicle $i$'s velocity and (controllable)
acceleration. The dynamics of vehicle $i$ are
\begin{equation}
\vspace*{0pt}\dot{x}_{i}(t)=v_{i}(t),\text{ \ \ }\dot{v}_{i}(t)=u_{i}(t)
\label{vehicle_dynamics}%
\end{equation}
The maneuvers carried out by vehicles $1,2,C$ are initiated at time $t_{0}$
and end at time $t_{f}$. 
We define
$d_{i}(v_{i}(t))$ to be the minimal safe distance between vehicle $i$ and the
one that precedes it in its lane, which in general depends on the vehicle's
current speed. The control input and speed are constrained as follows for all
$t\in\lbrack t_{0},t_{f}]$:
\begin{equation}
u_{i\text{min}}\leq u_{i}(t)\leq u_{i\text{max}},\text{ \ \ }v_{i\text{min}%
}\leq v_{i}(t)\leq v_{i\text{max}} \label{vehicle_constraint}%
\end{equation}
where $u_{i\text{max}}$, $u_{i\text{min}}$, $v_{i\text{max}}$, $v_{i\text{min}%
}$ are the maximal and minimal acceleration (respectively speed) limits. In
Fig. \ref{maneuver_process}, we control vehicles $1$, $2$ and $C$ to complete a
lane change maneuver while minimizing the maneuver time and the corresponding energy consumption. For each vehicle $i=1,2,C$ we formulate the following
optimization problem assuming that $x_{i}(0)$ and $v_{i}(0)$ are given:%

\begin{equation}
\begin{aligned}
J(t_{f};u_{i}(t))=&\min_{u_{i}(t)}\int_{0}^{t_{f}}[w_{t}+\\
&[w_{1,u}u_{1}%
^{2}(t)+w_{2,u}u_{2}^{2}(t)+w_{C,u}u_{C}^{2}(t)]]dt\\
\label{cost_function}
\end{aligned}
\end{equation}
\[
\begin{aligned} & \text{s.t. \ (\ref{vehicle_dynamics}), (\ref{vehicle_constraint}) and}\\ x_{1}(t)-x_{2}(t) & >d_{2}(v_{2}(t)),\text{ \ }t\in\lbrack0,t_{f}]\\ x_{U}(t)-x_{C}(t) & >d_{C}(v_{C}(t)),\text{ \ }t\in\lbrack0,t_{f}]\\ x_{1}(t_{f})-x_{C}(t_{f}) & >d_{C}(v_{C}(t_{f})),\text{ \ \ }x_{C}(t_{f})-x_{2}(t_{f})>d_{2}(v_{2}(t_{f})) \end{aligned}
\]
where $w_{t}$, $w_{u}$ are weights associated with the maneuver time $t_{f}$
and with a measure of the total energy expended. The two terms in the previous function need to be properly normalized and we set $w_{t}%
=\frac{\rho}{T_{\max}}$ and $w_{i,u}=\frac{1-\rho}{\text{max}\{u_\text{imax}%
^{2},u_\text{imin}^{2}\}}$, where $\rho\in\lbrack0,1]$ and $T_{\max}$ is a
prespecified upper bound on the maneuver time (e.g., $T_{\max}=l/\min\{v_\text{imin}\}$, $i=1,2,C,U$,
where $l$ is the distance to the next highway exit). Clearly, if $\rho=0$ this
problem reduces to an energy minimization problem and if $\rho=1$
it reduces to minimizing the maneuver time. The safe distance is defined as
$d_{i}(v_{i}(t))=\phi v_{i}(t)+\delta$ where $\phi$ is the headway time (the
general rule $\phi=1.8$ is usually adopted as in \cite{vogel2003comparison}).
As stated, the problem allows for a free terminal time $t_{f}$ and terminal
state constraints $x_{i}(t_{f})$, $v_{i}(t_{f})$. We will next specify the
terminal time $t_{f}$ as the solution of a minimization problem which allows
each vehicle to specify a desired \textquotedblleft aggressiveness
level\textquotedblright\ relative to the shortest possible maneuver time
subject to (\ref{vehicle_constraint}). Then, we will also specify $x_{i}%
(t_{f})$, $i=1,2,C$.

\section{Optimal Control Solution}

\textbf{Terminal time specification.} We begin by formulating the following
minimization problem based on which the maneuver terminal time $t_{f}$ is specified:%

\begin{equation}
\min_{t_{f}>0}\text{  }t_{f} \label{terminal time}%
\end{equation}%
\[
\begin{aligned}  \text{ s.t.  }x_1(0)&+v_1(0)t_f+0.5\alpha_1 u_{1\text{max}}t_f^2\\ -&x_C(0)-v_C(0)t_f-0.5\alpha_C u_{C\text{max}}t_f^2>d_C(v_C(t_f)) &\text{(4a)}\\ x_U(t_f)&-x_C(0)-v_C(0)t_f\\
-&0.5\alpha_C u_{C\text{min}}t_f^2>d_C(v_C(t_f))&\text{(4b)}\\ 
x_C(0)&+v_C(0)t_f+0.5\alpha_C u_{C\text{min}}t_f^2 \\ -&x_2(0)-v_2(0)t_f-0.5\alpha_2 u_{2\text{min}}t_f^2>d_2(v_2(t_f))&\text{(4c)}\\ \end{aligned}
\]
where $\alpha_{i}\in\lbrack0,1),i=1,2,C$ is an \textquotedblleft
aggressiveness coefficient\textquotedblright\ for vehicle $i$ which can be
preset by the driver. Observe that $[x_{i}(t_{0})+v_{i}(t_{0})t_{f}%
+0.5\alpha_{i}u_{i\text{max}}t_{f}^{2}]$ is the terminal position of $i$ under
control $\alpha_{i}u_{i\text{max}}$. To minimize $t_f$, vehicle $1$ should accelerate and vehicle $2$ decelerate so as to increase the gap between them in Fig. \ref{maneuver_process}. If $C$ accelerates, then (4a) ensures the safety constraint is still satisfied. If $C$ has to decelerate because it is constrained by $U$, then (4b) ensures that the safety constraint between $U$ and $C$ is satisfied and (4c) ensures that the safety constraint between $2$ and $C$ is also satisfied. As we will subsequently show, the optimal control of $C$ is either
always non-positive or always non-negative throughout $[0,t_{f}]$ so that either
the first or the last two constraints are relevant to it. Naturally, a
solution to \eqref{terminal time} may not exist, in which case we must iterate
on the values of $\alpha_{i}$ until one is possibly identified. If that is not
possible, then the maneuver is clearly aborted. If $t_{f}$ exists, we will
specify terminal position $x_{i}(t_{f})$ next and check the feasibility of
$(x_{i,f},t_{f})$ later in this section.

\textbf{Terminal position specifications.} Assuming a solution $t_{f}$ is
determined, we next seek to specify terminal vehicle positions $x_{i}(t_{f}),$
$i=1,2,C$, to be associated with problem \eqref{cost_function}. To do so, we
define%
\[
\Delta x_{i}(t_{f})=x_{i}(t_{f})-x_{i}(t_{0})-v_{i}(t_{0})t_{f}%
\]
which is the difference between the actual terminal position of $i$ and its
ideal terminal position under constant speed $v_{i}(t_{0})$; this is ideal
from the energy point of view in \eqref{terminal position}, since the energy
component is minimized when $u_{i}(t)=0$. Thus, the energy-optimal value is
$\Delta x_{i}(t_{f})=0$. We then seek terminal positions that minimize a
measure of deviating form these energy-optimal values over all three vehicles:%

\begin{equation}
\begin{aligned} \min_{x_i(t_f)>x_i(0),i=1,2,C}&\Delta x^2_C(t_f)+\Delta x^2_1(t_f)+\Delta x^2_2(t_f)\\ \text{s.t. }&\Delta x_i(t_f)=x_i(t_f)-x_i(0)-v_i(t_0)t_f\\ &x_1(t_f)-x_C(t_f)>\max\{d_C(v_C(t))\}\\ &x_C(t_f)-x_2(t_f)>\max\{d_2(v_2(t))\}\\ &x_U(t_f)-x_C(t_f)>\max\{d_C(v_C(t))\}\\ \end{aligned} \label{terminal position}%
\end{equation}
The max values in \eqref{terminal position} are assumed to be given by a
prespecified maximum inter-vehicle safe distance. However, as subsequently
shown in Theorem $1$, they actually turn out to be the known initial or
terminal values of $d_{2}(v_{2}(t))$ and $d_{C}(v_{C}(t))$. For example,
$\max\{d_{2}(v_{2}(t))\}=d_{2}(v_{2}(t_{0}))$ and $\max\{d_{C}(v_{C}%
(t))\}=d_{C}(v_{C}(t_{0})+u_{C\max}t_{f})$.

\textbf{Lemma $1$}: The solution $x_{i}^{\ast}(t_{f})$, $i=1,2,C$, to
(\ref{terminal position}) satisfies $\Delta x_{1}^{\ast}(t_{f})\geq0$ and
$\Delta x_{2}^{\ast}(t_{f})\leq0$.

\emph{Proof}\textbf{:} If $\Delta x_{1}^{\ast}(t_{f})<0$, then $\Delta
x_{1}(t_{f})=0$ is a better solution since it is feasible (the distance
between vehicles $1$, $2$ under $\Delta x_{1}(t_{f})=0$ is larger than under
$\Delta x_{1}^{\ast}(t_{f})<0$) and it is obvious that it yields a lower cost
in \eqref{terminal position} than the one with $\Delta x_{1}^{\ast}(t_{f})<0$
(the control is $u_{i}(t)=0$.) Therefore, we must have $\Delta x_{1}^{\ast
}(t_{f})\geq0$. The proof for $\Delta x_{2}^{\ast}(t_{f})\leq0$ is similar.
$\blacksquare$

\subsection{Optimal Control of Vehicles $1$ and $2$}

With the terminal time $t_{f}$ and longitudinal position $x_{i}(t_{f})$, $i=1,2$, set through (\ref{terminal time}) and
(\ref{terminal position}) respectively, the optimal control problems of
vehicles $i=1,2$ in \eqref{cost_function} become:%
\begin{equation}
\min_{u_{1}(t)}\int_{0}^{t_{f}}\frac{1}{2}u_{1}^{2}(t)dt\text{ \ \ \ s.t.
\ (\ref{vehicle_dynamics}), (\ref{vehicle_constraint}), }x_{1}(t_{f})=x_{1,f}
\label{OCP1}%
\end{equation}%
\begin{gather}
\min_{u_{2}(t)}\int_{0}^{t_{f}}\frac{1}{2}u_{2}^{2}(t)dt\text{ \ \ \ s.t.
\ (\ref{vehicle_dynamics}), (\ref{vehicle_constraint}), }x_{2}(t_{f})\leq
x_{2,f},\label{OCP2}\\
x_{1}(t)-x_{2}(t)>d_{2}(v_{2}(t)),\text{ \ }t\in\lbrack0,t_{f}]\nonumber
\end{gather}
where $x_{1,f}$ and $x_{2,f}$ are given above.
In (\ref{OCP2}), we use an inequality $x_{2}(t_{f})\leq x_{2,f}$ to describe
the terminal position constraint instead of the equality since it suffices for
the distance between the two vehicles to accommodate vehicle $C$ while at the
same time allowing for the cost under a control with $x_{2}(t_{f})<x_{2,f}$ to
be smaller than under a control with $x_{2}(t_{f})=x_{2,f}$. In (\ref{OCP1}),
there is no need to consider the case that $x_{1}(t_{f})>x_{1,f}$ since it is
clear that the optimal cost when $x_{1}(t_{f})=x_{1,f}$ is always smaller
compared to $x_{1}(t_{f})>x_{1,f}$. The next result establishes the fact that
the solution of these two problems involves vehicle 1 never decelerating and
vehicle 2 never accelerating.

\textbf{{Theorem $1$}} The optimal control in (\ref{OCP1}) is $u_{1}^{\ast
}(t)\geq0$ and the optimal control in (\ref{OCP2}) is $u_{2}^{\ast}(t)\leq0$.

\emph{Proof}:\textbf{ } First, by Lemma $1$, it is obvious that $u_{1}%
(t)\geq0$ is a feasible solution of (\ref{OCP1}) since $\Delta x_{1}^{\ast
}(t_{f})\geq0$ implies that $u_{1}(t)<0$ for all $t\in\lbrack t_{0},t_{f}]$ is
not feasible. The same applies to $u_{2}(t)\leq0$ being a feasible solution of
(\ref{OCP2}).

Starting with vehicle $1$, suppose that there exists some $[t_{1}%
,t_{2})\subset\lbrack0,t_{f}]$ in which the optimal solution satisfies
$u_{1}^{\ast}(t)<0$. We will show that there exists another control which
would lead to a smaller cost than $u_{1}^{\ast}(t)$. Consider a control
$u_{1}^{1}(t)$ defined so that $u_{1}^{1}(t)=u_{1}^{\ast}(t)\geq0$ for
$t\in\lbrack0,t_{1})\cup\lbrack t_{2},t_{f}]$, $u_{1}^{1}(t)=0$ for
$t\in\lbrack t_{1},t_{2})$. It is obvious that the cost of the control
$u_{1}^{1}(t)$ is lower than that of $u_{1}^{\ast}(t)$. However, we have
$x_{1}^{1}(t_{f})>x_{1}^{\ast}(t_{f})$ because $u_{1}^{1}(t)>u_{1}^{\ast}(t)$,
$t\in\lbrack t_{1},t_{2})$, thus violating the terminal condition in
(\ref{OCP1}). Therefore, we construct another control $u_{1}^{2}(t)$, a
variant of $u_{1}^{1}(t)$ which is feasible, as follows. Define
\begin{equation}
g_{1}(t)=x_{1}^{1}(t)+v_{1}^{1}(t)(t_{f}-t) \label{g1}%
\end{equation}
and observe that $g_{1}(t)$ is a continuous function of $t$ since $x_{1}%
^{1}(t)$ and $v_{1}^{1}(t)$ are continuous. Because $g_{1}(0)={x}_{1}%
^{1}(0)+v_{1}^{1}(0)t_{f}<x_{1}^{\ast}(t_{f})$ (by Lemma $1$) and $g_{1}%
(t_{f})=x_{1}^{1}(t_{f})\geq x_{1}^{\ast}(t_{f})$, there exists some $t_{m}%
\in\lbrack0,t_{f}]$ such that $g_{1}(t_{m})=x_{1}^{\ast}(t_{f})$. We now
define a control $u_{1}^{2}(t)$ such that $u_{1}^{2}(t)=u_{1}^{1}(t)\geq0$ for
$t\in\lbrack0,t_{m})$, $u_{1}^{2}(t)=0$ for $t\in\lbrack t_{m},t_{f}]$. It
follows that $x_{1}^{2}(t_{m})=x_{1}^{1}(t_{m})$, $v_{1}^{2}(t_{m})=v_{1}%
^{1}(t_{m})$ and $x_{1}^{2}(t_{f})=x_{1}^{1}(t_{m})+v_{1}^{1}(t_{m}%
)(t_{f}-t_{m})$ which implies that $x_{1}^{2}(t_{f})=g_{1}(t_{m})$ from
\eqref{g1}. Thus, the terminal position constraint is not violated under
$u_{1}^{2}(t)$. Based on the definitions of $u_{1}^{2}(t)$ and $u_{1}^{1}(t)$,
it is obvious that $u_{1}^{2}(t)$ does not violate the acceleration
constraints in (\ref{vehicle_constraint}). Next, we show that the velocity
constraints in (\ref{vehicle_constraint}) are also not violated. Assume that
for some $t_{n}$, $v_{1}^{1}(t_{n})=v_{\text{1max}}$ initiating an arc where
the velocity is $v_{1}^{1}(t)=v_{\text{1max}}$. There are two cases:

\textbf{(a)} If $t_{n}\geq t_{m}$, we have $v_{1}^{1}(t_{m})\leq v_{1}%
^{1}(t_{n})$ because $u_{1}^{1}(t)\geq0$ for all $t\in\lbrack0,t_{f}]$. Based
on the definition of $u_{1}^{2}(t)$, the maximal speed under the control
$u_{1}^{2}(t)$ is $v_{1}^{1}(t_{m})$ and the velocity constraint is,
therefore, inactive.

\textbf{(b)} If $t_{n}<t_{m}$, we have $v_{1}^{1}(t_{m})>v_{1}^{1}(t_{n})$.
Taking the time derivative of $g_{1}(t)$, we get $\dot{g}_{1}(t)=u_{1}%
^{1}(t)(t_{f}-t)\geq0$. It follows that
\begin{equation}
g_{1}(t_{n})<g_{1}(t_{m})=x_{1}^{\ast}(t_{f}) \label{g1inequality}%
\end{equation}
where the equality follows from the definition of $t_{m}$ above. Then, let us
construct a new control $u_{1}^{3}(t)$ such that $u_{1}^{3}(t)=u_{1}^{\ast
}(t)\geq0$ for $t\in\lbrack0,t_{1})\cup\lbrack t_{2},t_{n})$, $u_{1}^{3}(t)=0$
for $t\in\lbrack t_{1},t_{2})\cup\lbrack t_{n},t_{f}]$, where $t_{n}\geq
t_{1}$ because $v_{1}^{3}(t)<v_\text{1max}$ for $t<t_{1}$ based on the feasibility
of $u_{1}^{\ast}(t)$. Moreover, if $t_{1}<t_{n}<t_{2}$, we define $u_{1}%
^{3}(t)=0$ for $t>t_{1}$, i.e., $u_{1}^{3}(t)=u_{1}^{\ast}(t)\geq0$ for
$t\in\lbrack0,t_{1})$, $u_{1}^{3}(t)=0$ for $t\in\lbrack t_{1},t_{f}]$. Based
on the definition of $u_{1}^{3}(t)$, we have $u_{1}^{3}(t)=u_{1}^{1}(t)$ for
$t\in\lbrack0,t_{n})$. Therefore, $x_{1}^{3}(t_{n})=x_{1}^{1}(t_{n})$ and
$v_{1}^{3}(t_{n})=v_{1}^{1}(t_{n})$ so that (\ref{g1inequality}) holds under
$u_{1}^{3}(t)$. When $t\geq t_{n}$, we have $u_{1}^{3}(t)=0$, therefore,
$x_{1}^{3}(t_{f})=x_{1}^{1}(t_{n})+v_{1}^{1}(t_{n})(t_{f}-t_{n})=g_{1}(t_{n})$
from \eqref{g1}. Since $v_{1}^{3}(t)=v_{1\text{max}}$ for $t\in\lbrack t_{n},t_{f}]$
and $x_{1}^{3}(t)\geq x_{1}^{\ast}(t)$ for $t\in\lbrack0,t_{n})$, it is clear
that $x_{1}^{3}(t_{f})\geq x_{1}^{\ast}(t_{f})$. However, since $g_{1}%
(t_{n})=x_{1}^{3}(t_{f})\geq x_{1}^{\ast}(t_{f})$, this contradicts
\eqref{g1inequality}. We conclude that $t_{n}<t_{m}$ is not possible. In
summary, we have shown that the velocity constraint is inactive for control
$u_{1}^{2}(t)$. Therefore, $u_{1}^{2}(t)$ is feasible and results in a lower
cost in (\ref{OCP1}) than $u_{1}^{\ast}(t)$ since it includes a trajectory arc
over which $u_{1}^{2}(t)=0$. This contradicts the optimality of $u_{1}^{\ast
}(t)$ and we conclude that the optimal control cannot contain any interval
over which $u_{1}^{\ast}(t)<0$.

Next, consider vehicle 2 and suppose that there exists some $[t_{1}%
,t_{2})\subset\lbrack0,t_{f}]$ in which the optimal solution satisfies
{$u_{2}^{\ast}(t)>0$}. Consider a control $u_{2}^{1}(t)$ defined so that
{$u_{2}^{1}(t)=u_{2}^{\ast}(t)\leq0$ }for $t\in\lbrack0,t_{1})\cup\lbrack
t_{2},t_{f}]$, $u_{1}^{1}(t)=0$ for $t\in\lbrack t_{1},t_{2})$. {It is clear
that the cost under $u_{2}^{1}(t)$ is} lower than that of $u_{2}^{\ast}(t)$
and that {the acceleration constraint in (\ref{vehicle_constraint}) is
inactive for $u_{2}^{1}(t)$. Furthermore, it is obvious that $x_{2}^{\ast
}(t)\geq x_{2}^{1}(t)$ and $v_{2}^{\ast}(t)\geq v_{2}^{1}(t)$ for $t\in
\lbrack0,t_{f}]$. Therefore, the terminal position inequality in (\ref{OCP2})
is not violated. Based on the definition of the safety distance constraint,
$d_{2}(v_{2}(t))$}$=\phi v_{2}(t)+\delta${ is monotonically increasing in
$v_{2}(t)$. } {Therefore, we conclude that the safety constraint under
$u_{2}^{1}(t)$ will not be violated, since $u_{2}^{\ast}(t)$ is feasible and
$x_{2}^{\ast}(t)\geq x_{2}^{1}(t)$, $v_{2}^{\ast}(t)\geq v_{2}^{1}(t)$.
Finally, we consider the speed constraint in (\ref{vehicle_constraint}) which
may be active under $u_{2}^{1}(t)$. There are two cases:}

{\textbf{(a)} If $v_{2}^{1}(t_{f})>v_{2\text{min}}$, the speed constraint is
inactive under $u_{2}^{1}(t)$ over all }$t\in\lbrack0,t_{f}]$ {and $u_{2}%
^{1}(t)$ is a feasible solution which results in }a lower cost in (\ref{OCP2})
than $u_{2}^{\ast}(t)$ since it includes a trajectory arc over which
$u_{2}^{2}(t)=0$.

{\textbf{(b)} If $v_{2}^{1}(t_{f})\leq v_{2\text{min}}$, there must exist some
$t_{n}\in\lbrack t_{1},t_{f}]$ such that $v_{2}^{1}(t_{m})=v_{\text{2min}}$.
Let us construct a new control ${u}_{2}^{2}(t)$ as follows: ${u}_{2}%
^{2}(t)=u_{2}^{1}(t)\leq0$ for }$t\in\lbrack0,t_{m})$, {${u}_{2}^{2}(t)=0$ for
}$t\in\lbrack t_{m},t_{f}]${. For $t\in\lbrack0,t_{m})$, it is obvious that
$x_{2}^{\ast}(t)\geq x_{2}^{2}(t)$ and $v_{2}^{\ast}(t)\geq v_{2}^{2}(t)$
based on the definition of $u_{2}^{1}(t)$. For $t\in\lbrack t_{m},t_{f}]$,
vehicle $2$ moves at the minimal speed $v_{2\text{min}}$ under $u_{2}^{2}(t)$,
therefore, $x_{2}^{\ast}(t_{f})\geq x_{2}^{2}(t_{f})$, that is, the terminal
position inequality is satisfied. Also, it is obvious that the acceleration
and the speed constraints are not violated over $[0,t_{f}]$. }{Finally, we
have shown that $u_{2}^{1}(t)$ does not violate the safety constraint. Based
on the same argument, it is straightforward to show that $u_{2}^{2}(t)$ will
not violate this constraint, since $x_{2}^{2}(t)\leq x_{2}^{\ast}(t)$ and
$v_{2}^{2}(t)\leq v_{2}^{\ast}(t)$. Therefore, $u_{2}^{2}(t)$ is feasible in
(\ref{OCP2}) and the corresponding cost is lower than that of $u_{2}^{\ast
}(t)$ because the trajectory segment with $u_{2}^{2}(t)=0$ contributes to zero
cost. We conclude that the optimal control $u_{2}^{\ast}(t)$ cannot contain
any time interval with $u_{2}^{\ast}(t)>0$}. $\blacksquare$

Based on Theorem $1$, in addition to showing that vehicle 1 never decelerates
and vehicle 2 never accelerates, we also eliminate the safe distance
constraint in \eqref{OCP2} since the distance between the vehicles will
increase in the course of the maneuver and the last two safety constraints in
\eqref{cost_function} ensure that this distance is eventually large enough to
accommodate the length of vehicle $C$. Thus, (\ref{OCP2}) becomes%
\begin{equation}
\min_{u_{2}(t)}\int_{0}^{t_{f}}\frac{1}{2}u_{2}^{2}(t)dt\text{ \ \ \ s.t.
\ (\ref{vehicle_dynamics}), (\ref{vehicle_constraint}), }x_{2}(t_{f})=x_{2,f}
\label{OCP2_rev}%
\end{equation}



\begin{figure}[pt]
\centering
\includegraphics[scale=0.55]{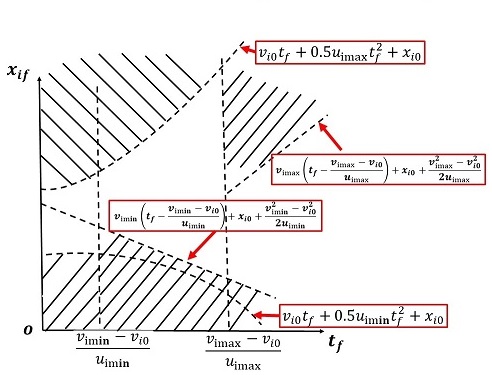} \vspace*{-\baselineskip}
\caption{The feasible state set of controllable vehicles in the left lane.}%
\label{feasibleset}%
\end{figure}

\textbf{Feasible terminal state set. }The constraints in
(\ref{vehicle_constraint}) limit the sets of feasible terminal conditions
$(x_{i,f},t_{f})$, $i=1,2,C$ as shown in Fig. \ref{feasibleset} where the
feasible set is the unshaded area defined as follows for each $i=1,2,C$: $(i)$
Vehicle $i$ cannot reach $x_{i,f}$ under its maximal acceleration if $u_{i\max
}t_{f}+v_{i,0}\leq v_{i\max}$ and $v_{i,0}t_{f}+0.5u_{i\max}t_{f}^{2}%
<x_{i,f}-x_{i,0}$. $(ii)$ Vehicle $i$ cannot reach $x_{i,f}$ under its maximal
acceleration after attaining its maximal velocity if $u_{i\text{max}}%
t_{f}+v_{i,0}>v_{i\text{max}}$ and $v_{i\text{max}}(t_{f}-\frac{v_{i\text{max}%
}-v_{i,0}}{u_{i\text{max}}})<x_{i,f}-x_{i,0}-\frac{v_{i\text{max}}^{2}-v_{i,0}%
^{2}}{2u_{i\text{max}}}$. $(iii)$ Vehicle $i$ exceeds $x_{i,f}$ under the
minimal acceleration if $u_{i\text{min}}t_{f}+v_{i,0}\geq v_{i\text{min}}$ and
$v_{i,0}t_{f}+0.5u_{i\text{min}}t_{f}^{2}>x_{i,f}-x_{i,0}$. $(iv)$ Vehicle $i$
exceeds $x_{i,f}$ under the minimal acceleration after attaining its minimal
velocity if $u_{i\text{min}}t_{f}+v_{i,0}<v_{i\text{min}}$ and $v_{i\text{min}%
}(t_{f}-\frac{v_{i\text{min}}-v_{i,0}}{u_{i\text{min}}})>x_{i,f}-x_{i,0}%
-\frac{v_{i\text{min}}^{2}-v_{i,0}^{2}}{2u_{i\text{min}}}$. In addition,
vehicle $C$ must also satisfy a safety distance constraint with respect to
vehicle $U$, hence if $x_{C,f}>x_{U}(0)+v_{U}t_{f}-d_{C}(v_{2}(t_{f}))$, there
is no feasible solution.

Note that if an optimal $t_{f}$ is determined in \eqref{terminal time} and the
solution of \eqref{terminal position} guarantees that $x_{i}(t_{f})$,
$i=1,2,C$, do not violate the safety constraints, $(x_{i,f},t_{f})$ is expected
to be feasible. However, if $(x_{i,f},t_{f})$ is infeasible for vehicle $i$,
then the following algorithm is used to find a feasible such pair:

\textbf{Algorithm 1:}

(1) $t_{f}$ is updated using $t_{f}=\beta t_{f}$, $\beta>1$.

(2) With updated $t_{f}$, \eqref{terminal position} is re-solved to obtain new
$x_{i,f}$.

(3) If $(x_{i,f},t_{f})$ is feasible in Fig. \ref{feasibleset}, stop; else
return to step (1) with a higher value of $\beta$.

In the above, the coefficient $\beta$ is used to relax the maneuver time
$t_{f}$ so as to accommodate one or more of the constraints in Fig.
\ref{feasibleset} until a feasible $(x_{i,f},t_{f})$ is identified.

\textbf{Solution of problem (\ref{OCP1})}. We can now proceed to derive an
explicit solution for (\ref{OCP1}) taking advantage of Theorem $1$. We begin
by writing the Hamiltionian and associated Lagrangian functions for
\eqref{OCP1}:
\begin{equation}
H(v_{1},u_{1},\lambda)=\frac{1}{2}u_{1}^{2}(t)+\lambda_{x}(t)v_{1}(t)+\lambda
_{v}(t)u_{1}(t) \label{hamiltion}%
\end{equation}%
\begin{equation}
\begin{aligned} L(v_1,u_1,\lambda,\eta)=&H(v,u,\lambda)+\eta_{1}(t)(u_\text{1min}-u_1(t))\\ &+\eta_{2}(t)(u_1(t)-u_\text{1max})+\eta_{3}(t)(v_\text{1min}-v_1(t))\\ &+\eta_{4}(t)(v_1(t)-v_\text{1max})\\ \end{aligned} \label{L_Cost1}%
\end{equation}
where $\lambda(t)=[\lambda_{v}(t),\lambda_{x}(t)]^{T}$ and $\eta=[\eta
_{1}(t),...,\eta_{4}(t)]^{T}$. In view of\ Theorem 1, i.e., $u_{1}^{\ast
}(t)\geq0$\textbf{, }\eqref{L_Cost1} reduces to
\begin{equation}
\begin{aligned} L(v_1,u_1,\lambda,\eta)=&\frac{1}{2}u_1^{2}(t)+\lambda_{x}(t)v_1(t)+\lambda_{v}(t)u_1(t)\\&+\eta_{2}(t)(u_1(t)-u_\text{1max})+\eta_{3}(t)(v_\text{1min}-v_1(t))\\
&+\eta_{4}(t)(v_1(t)-v_\text{1max})\\ \end{aligned}
\end{equation}
The explicit solution of (\ref{OCP1}) is given next.

\textbf{{Theorem $2$}} Let $x_{1}^{\ast}(t)$, $v_{1}^{\ast}(t)$, $u_{1}^{\ast
}(t)$ be a solution of (\ref{OCP1}). Then,
\begin{equation}
u_{1}^{\ast}(t)=\arg\min_{0\leq u_{1}\leq u_{1\text{max}}}\frac{1}{2}[u_{1}%
^{2}+\dfrac{u_{1}^{\ast}(t_{0})^{2}(t-\tau)u_{1}}{v_{1,0}-v_{1}^{\ast}%
(t_{f})+(\tau-t_{0})u_{1}^{\ast}(t_{0})}] \label{OCP1_solution}%
\end{equation}
where $\tau$ is the first time that $v_{1}^{\ast}(\tau)=v_{1\text{max}}$ and
$\tau=t_{f}$ if $v_{1\text{max}}$ is never reached.

\emph{Proof}:\textbf{ } Problem (\ref{OCP1}) is of the same form as the fixed
terminal time optimal control Problem 3 in \cite{meng2018optimal} whose
solution when $u_{1}^{\ast}(t)\geq0$ is given in Theorem 2 of
\cite{meng2018optimal} and is therefore omitted. By Pontryagin's principle
applied to (\ref{hamiltion}), $u_{1}^{\ast}(t)=$min$\{u_{1\max},{-\lambda
_{v}(t)}\}$ and the key parts of the proof in \cite{meng2018optimal} are
showing that $\eta_{3}(t)=0$ and that $\lambda_{v}(t)$ is continuous for all
$t\in\lbrack0,t_{f}]$. $\blacksquare$

Furthermore, following a derivation similar to that in \cite{meng2018optimal}
we can obtain the optimal cost $J_{1}^{\ast}(t_{f})$ in (\ref{OCP1}) based on
several cases depending on the initial acceleration $u_{1,0}^{\ast}$ and the
terminal velocity $v_{1}^{\ast}(t_{f})$ which can be explicitly evaluated as
in \cite{meng2018optimal}. The final optimal cost is the minimal among all
possible values obtained.

\textbf{Case I: }$u_{1,0}^{\ast}=u_{1\text{max}}$ and $\dot{u}_{1}^{\ast
}(t)=0$. If $t_{f}<\frac{v_{1\text{max}}-v_{1,0}}{u_{1\text{max}}}$, then
$u_{1}^{\ast}(t)=u_{1\text{max}}$ for all $t\in\lbrack0,t_{f}]$. Otherwise,
when $v_{1}(t)=v_{1\text{max}}$, the control switches to $u_{1}^{\ast}(t)=0$.
Therefore,
\begin{equation}
J_{1}^{\ast}(t_{f})=\left\{  \vspace*{0pt}%
\begin{array}
[c]{l}%
\frac{1}{2}u_{1\text{max}}(v_{1\text{max}}-v_{1,0})\\
\frac{1}{2}u_{1\text{max}}^{2}(t_{f}-t_{0})\\
\end{array}
\right.
\begin{array}
[c]{l}%
\text{if }t_{f}\geq\frac{v_{1\text{max}}-v_{1,0}}{u_{1\text{max}}}\\
\text{otherwise }\\
\end{array}
\label{case1}%
\end{equation}

\textbf{Case II: }$u_{1,0}^{\ast}=u_{1\text{max}}$ and ${v}_{1}^{\ast}%
(t_{f})=v_{1\text{max}}$. We define $t_{1}$ as the time that $u_{1}^{\ast}(t)$
begins to decrease and $\tau$ as the first time that $u_{1}^{\ast}(\tau)=0$.
Thus, $u_{1}^{\ast}(t)$ is a piecewise linear function of time $t$ and
(following calculations similar to those in \cite{meng2018optimal}):
\begin{equation}
J_{1}^{\ast}(t_{f})=\frac{1}{2}(t_{1}-t_{0})u_{1\text{max}}^{2}+\frac{1}%
{24}\frac{u_{1\text{max}}^{4}(\tau-t_{1})^{3}}{[v_{1,0}-v_{1\text{max}}%
+(\tau-t_{0})u_{1\text{max}}]^{2}}%
\end{equation}
Using similar calculations, we summarize below the remaining three cases:%
\begin{align*}
\text{\textbf{Case III}}  &  \mathbf{: \ \ }\text{}%
\begin{array}
[c]{c}%
u_{1,0}^{\ast}=u_{1\text{max}}\\
{v}_{1}^{\ast}(t_{f})<v_{1\text{max}}%
\end{array}
\text{ \ \ }J_{1}^{\ast}(t_{f})=\frac{1}{2}\frac{u_{1\text{max}}^{2}%
(t_{f}+2t_{1}-3t_{0})}{3}\\
\text{\textbf{Case IV}}  &  \mathbf{: \ \ }\text{}%
\begin{array}
[c]{c}%
u_{1,0}^{\ast}<u_{1\text{max}}\\
{v}_{1}^{\ast}(t_{f})=v_{1\text{max}}%
\end{array}
\text{ \ \ }J_{1}^{\ast}(t_{f})=\frac{2}{3}\frac{(v_{1\text{max}}-v_{1,0}%
)^{2}}{\tau-t_{0}}\\
\text{\textbf{Case V}}  &  \mathbf{: \ \ }\text{}%
\begin{array}
[c]{c}%
u_{1,0}^{\ast}<u_{1\text{max}}\\
{v}_{1}^{\ast}(t_{f})<v_{1\text{max}}%
\end{array}
\text{ \ \ }J_{1}^{\ast}(t_{f})=\frac{3}{2}\frac{[x_{1,f}-v_{1,0}(t_{f}%
-t_{0})]^{2}}{(t_{f}-t_{0})^{3}}%
\end{align*}

\textbf{Solution of problem }(\ref{OCP2_rev}). Similar to the solution of
(\ref{OCP1}), we can derive an explicit solution for (\ref{OCP2_rev}) taking
advantage of Theorem $1$ and obtain the following result.

\textbf{{Theorem $3$}} Let $x_{2}^{\ast}(t)$, $v_{2}^{\ast}(t)$, $u_{2}^{\ast
}(t)$ be a solution of (\ref{OCP2_rev}). Then,
\begin{equation}
u_{2}^{\ast}(t)=\arg\min_{u_{2\text{min}}\leq u_{2}\leq0}\frac{1}{2}[u_{2}%
^{2}+\dfrac{u_{2}^{\ast}(t_{0})^{2}(\tau-t)u_{2}}{v_{2}^{\ast}(t_{f}%
)-v_{2,0}-(\tau-t_{0})u_{2}^{\ast}(t_{0})}]%
\end{equation}
where $\tau$ is the first time that $v_{2}^{\ast}(\tau)=v_{2\text{min}}$ and
$\tau=t_{f}$ if $v_{2\text{min}}$ is never reached.

\emph{Proof}:\textbf{ } Problem (\ref{OCP2}) is also of the same form as the
fixed terminal time optimal control Problem 3 in \cite{meng2018optimal} whose
solution when $u_{2}^{\ast}(t)\leq0$ is given in Theorem 3 of
\cite{meng2018optimal} and is therefore omitted. $\blacksquare$

We can also obtain the optimal cost $J_{2}^{\ast}(t_{f})$ in (\ref{OCP2})
based on several cases depending on the initial acceleration $u_{2,0}^{\ast}$
and the terminal velocity $v_{2}^{\ast}(t_{f})$ which can be explicitly
evaluated as in \cite{meng2018optimal}. In what follows, we define $t_{1}$ as
the time that $u_{2}^{\ast}(t)$ begins to increase and $\tau$ as the first
time that $u_{2}^{\ast}(\tau)=0$.%
\begin{align*}
\text{\textbf{Case I}}\mathbf{:\ \ }  &  \text{{}}%
\begin{array}
[c]{c}%
u_{2,0}^{\ast}=u_{2\text{min}}\\
{v}_{2}^{\ast}(t_{f})=v_{2\text{min}}%
\end{array}
\text{ \ \ }%
\begin{array}
[c]{c}%
J_{2}^{\ast}(t_{f})=\frac{1}{2}(t_{1}-t_{0})u_{2\text{min}}^{2}+\\
\frac{1}{24}\frac{u_{2\text{min}}^{4}(\tau-t_{1})^{3}}{[v_{2,0}-v_{2\text{min}%
}+(\tau-t_{0})u_{2\text{min}}]^{2}}%
\end{array}
\\
\text{\textbf{Case II}}\mathbf{:\ \ }  &  \text{{}}%
\begin{array}
[c]{c}%
u_{2,0}^{\ast}=u_{2\text{min}}\\
{v}_{2}^{\ast}(t_{f})>v_{2\text{min}}%
\end{array}
\text{ \ \ }J_{2}^{\ast}(t_{f})=\frac{u_{2\text{min}}^{2}(t_{f}+2t_{1}%
-3t_{0})}{6}\\
\text{\textbf{Case III}}\mathbf{:\ \ }  &  \text{{}}%
\begin{array}
[c]{c}%
u_{2,0}^{\ast}>u_{2\text{min}}\\
{v}_{2}^{\ast}(t_{f})=v_{2\text{min}}%
\end{array}
\text{ \ \ }J_{2}^{\ast}(t_{f})=\frac{2}{3}\frac{(v_{2\text{min}}-v_{2,0}%
)^{2}}{\tau-t_{0}}\\
\text{\textbf{Case IV}}\mathbf{:\ \ }  &  \text{{}}%
\begin{array}
[c]{c}%
u_{2,0}^{\ast}>u_{2\text{min}}\\
{v}_{2}^{\ast}(t_{f})=v_{2\text{min}}%
\end{array}
\text{ \ \ }J_{2}^{\ast}(t_{f})=\frac{3}{2}\frac{[x_{2,f}-v_{2,0}(t_{f}%
-t_{0})]^{2}}{(t_{f}-t_{0})^{3}}
\end{align*}

\subsection{Optimal Control of Vehicle $C$}

Unlike (\ref{OCP1}) and (\ref{OCP2_rev}), deriving the optimal control of
vehicle $C$ as in Fig. \ref{maneuver_process} is more challenging. First,
since we need to keep a safe distance between vehicles $C$ and $U$, a
constraint $x_{U}(0)+v_{U}t-x_{C}(t)>d_{C}(v_{C}(t))$ must hold for all
$t\in\lbrack0,t_{f}]$. The resulting problem formulation is:
\begin{equation}
\begin{aligned} \min_{u_C(t)}& \int_{0}^{t_f}\frac{1}{2}u_{C}^{2}(t)dt\\ \text{s.t. }& \eqref{vehicle_dynamics},\text{ }\eqref{vehicle_constraint},\text{ }x_{C}(t_f)=x_{C,f},\text{ }t\in[0,t_f]\\ &x_U(0)+v_Ut-x_C(t)>d_C(v_C(t))\\ \end{aligned} \label{cost2}%
\end{equation}
\newline in which $d_{C}(v_{C}(t))$ is time-varying. To simplify
\eqref{cost2}, we use $d_{C}\equiv$max$\{d_{C}(v_{C}(t))\}$ instead of
$d_{C}(v_{C}(t))$, which is a more conservative constraint still ensuring that
the original one is not violated (the problem with $d_{C}(v_{C}(t))=\phi
v_{C}(t)+\delta$ can still be solved at the expense of added complexity and is the
subject of ongoing research).

The Hamiltonian for (\ref{cost2}) with the constraints adjoined yields the
Lagrangian
\begin{gather}
L(x_{C},v_{C},u_{C},\lambda,\eta)=\frac{1}{2}u_{C}^{2}(t)+\lambda_{x}%
(t)v_{C}(t)+\lambda_{v}(t)u_{C}(t)\label{Lagrangian_3}\\
+\eta_{1}(t)(u_{C}(t)-u_{C\max})+\eta_{2}(t)(u_{C\min}-u_{C}(t))\nonumber\\
+\eta
_{3}(t)(v_{C}(t)-v_{C\max})
+\eta_{4}(t)(v_{C\min}-v_{C}(t))\nonumber\\
+\eta_{5}(t)(x_{C}(t)-x_{U}(0)-v_{U}%
(0)t+d_{C})\nonumber
\end{gather}
with $\lambda(t)=[\lambda_{v}(t),\lambda_{x}(t)]^{T}$ and $\eta=[\eta
_{1}(t),...,\eta_{5}(t)]^{T}$, $t\in\lbrack0,t_{f}]$. Based on Pontryagin's
principle, we have
\begin{equation}
u_{C}^{\ast}(t)=\left\{  \vspace*{0pt}%
\begin{array}
[c]{l}%
-\lambda_{v}(t)\\
u_{C\text{min}}\\
u_{C\text{max}}\\
\end{array}
\right.
\begin{array}
[c]{l}%
\text{if }u_{C\text{min}}\leq-\lambda_{v}(t)\leq u_{C\text{max}}\\
\text{if }-\lambda_{v}(t)<u_{C\text{min}}\\
\text{if }-\lambda_{v}(t)>u_{C\text{max}}%
\end{array}
\label{u_dynamics}%
\end{equation}
when none of the constraints is active along an optimal trajectory. In order
to account for the constraints becoming active, we identify several cases
depending on the terminal states of vehicles $U$ and $C$. Let us define
$\bar{x}_{C}(t_{f})$ to be the terminal position of $C$ if $u_{C}(t)=0$ for
all $t\in\lbrack0,t_{f}]$. The relationship between $\bar{x}_{C}(t_{f})$ and
${x}_{C}(t_{f})$ is critical. In particular, if $\bar{x}_{C}(t_{f})<{x}%
_{C}(t_{f})$, vehicle $C$ must accelerate in order satisfy the terminal
position constraint. Otherwise, $C$ must decelerate. Also critical is the
value of $x_{U}(t_{f})-d_{C}$, i.e., the upper bound of the safe terminal
position of $C$. In addition, during the entire maneuver process, we require
that ${x}_{C}(t)\leq x_{U}(t)-d_{C}$.

We begin with the $3!$ cases for ordering ${x}_{C}(t_{f})$, $\bar{x}_{C}%
(t_{f})$ and $x_{U}(t_{f})-d_{C}$. Fortunately, we can exclude several
cases as infeasible because ${x}_{C}(t_{f})\leq x_{U}(t_{f})-d_{C}$ is a
necessary condition to have feasible solutions. This leaves three remaining
cases as follows.

\emph{Case 1}: $\bar{x}_{C}(t_{f})<x_{C}(t_{f})<x_{U}(t_{f})-d_{C}$.

\emph{Case 2}: $x_{C}(t_{f})<\bar{x}_{C}(t_{f})<x_{U}(t_{f})-d_{C}$.

\emph{Case 3}: ${x}_{C}(t_{f})<x_{U}(t_{f})-d_{C}<\bar{x}_{C}(t_{f})$.

\begin{figure*}[h]
\centering
\includegraphics[width=.3\textwidth]{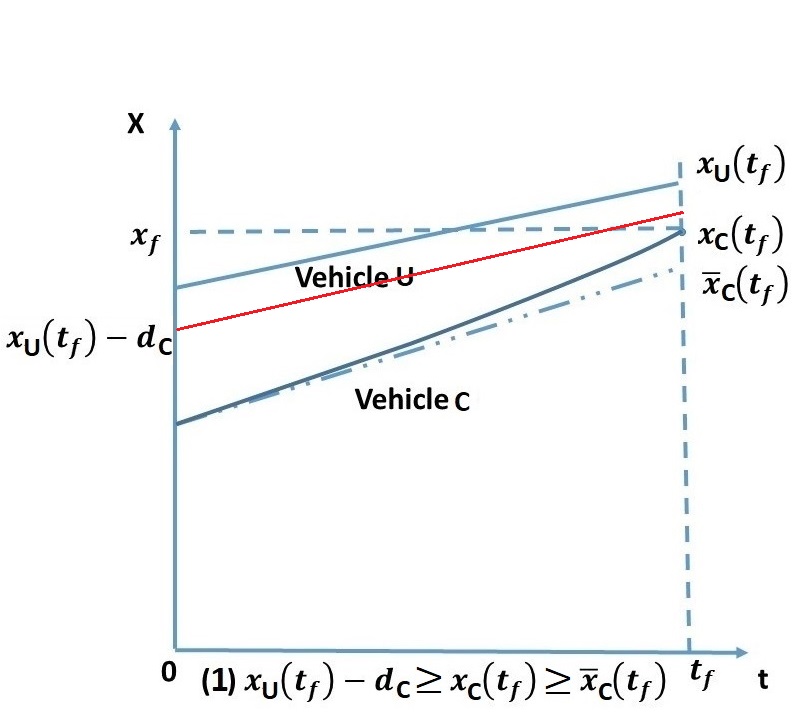}
\includegraphics[width=.3\textwidth]{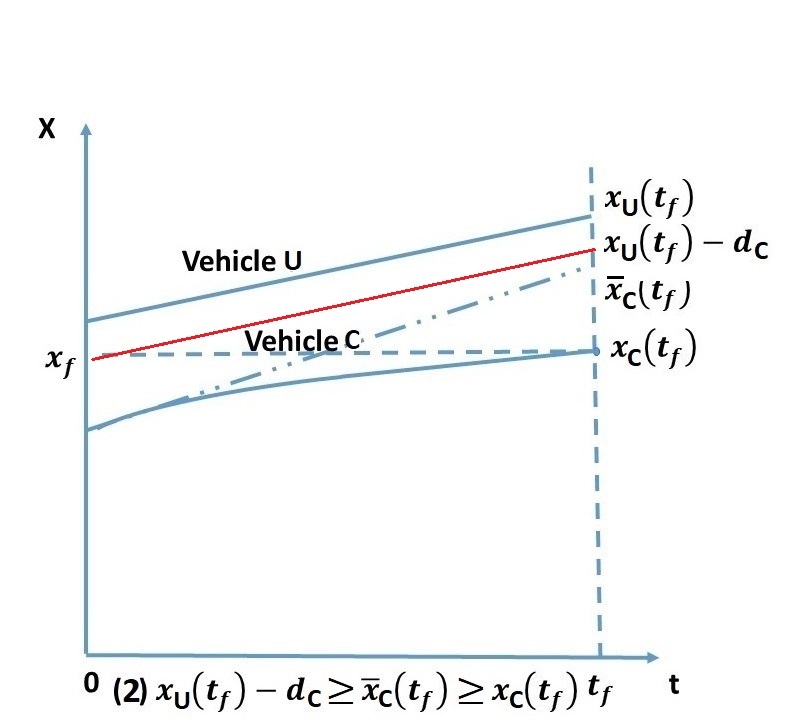}
\includegraphics[width=.3\textwidth]{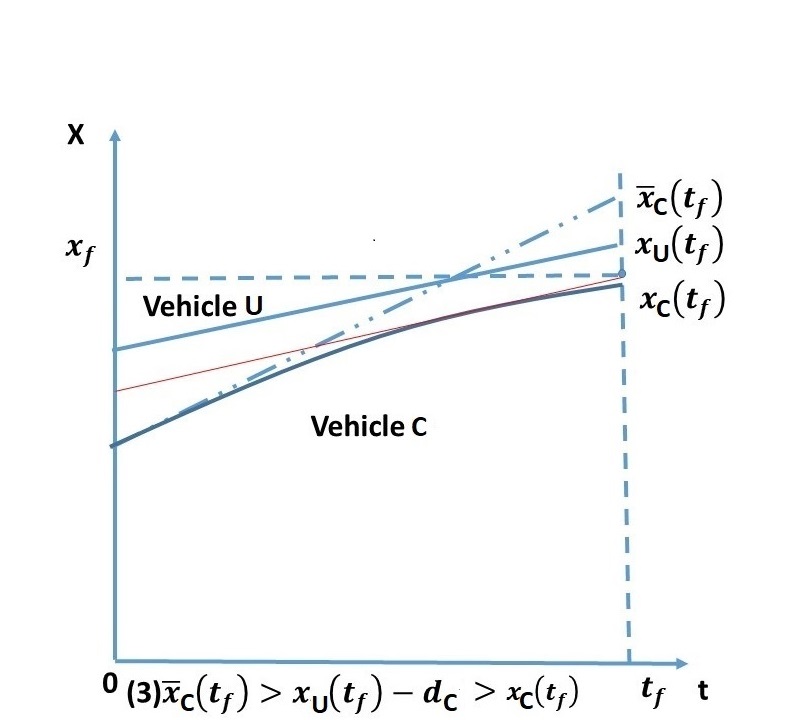}
\caption{The three feasible cases for the optimal maneuver of vehicle $C$.}%
\label{case_study}%
\end{figure*}

These are visualized in Fig. \ref{case_study}. The following results provide
structural properties of the optimal solution (\ref{u_dynamics}) depending on
which case applies.

\textbf{Lemma $2$}: If $x_{U}(0)+v_{U}(0)t-x_{C}(t)=d_{C}$, then
$v_{C}(t)=v_{U}(0)$, $t\in\lbrack0,t_{f}]$.

\emph{Proof}\textbf{:} Assume that at time $t_{k}$, we have $x_{U}%
(0)+v_{U}(0)t_{k}-x_{C}(t_{k})=d_{C}$ and define $f(t)=x_{U}(0)+v_{U}%
(0)t-x_{C}(t)$. Using a contradiction argument, if $v_{C}(t)\neq v_{U}(0)$
there are two cases: $(i)$ If $v_{C}(t_{k})>v_{U}(0)$, since $f^{^{\prime}%
}(t_{k})=v_{U}(0)-v_{C}(t_{k})<0$, we have $f(t_{k}^{-})>f(t_{k})$, which
implies $x_{U}(0)+v_{U}(0)t_{k}^{-}-x_{C}(t_{k}^{-})>d_{C}$, therefore, the
safety constraint is violated at $t_{k}^{-}$. $(ii)$ If $v_{C}(t_{k}%
)<v_{U}(0)$, since $f^{^{\prime}}(t_{k})=v_{U}(0)-v_{C}(t_{k})>0$, we have
$f(t_{k}^{+})>f(t_{k})$, which implies $x_{U}(0)+v_{U}(0)t_{k}^{+}-x_{C}%
(t_{k}^{+})>d_{C}$, therefore, the safety constraint is violated at $t_{k}%
^{+}$. We conclude that $v_{C}(t_{k})=v_{U}$ which completes the proof.
$\ \blacksquare$

\textbf{Theorem $4$} [\emph{Case 1} in Fig. \ref{case_study}]: If $\bar{x}%
_{C}(t_{f})<x_{C}(t_{f})<x_{U}(t_{f})-d_{C}$, then $u_{C}^{\ast}(t)\geq0$ and
$\eta_{5}^{\ast}(t)=0$.

\emph{Proof}\textbf{: }The condition $x_{C}(t_{f})>\bar{x}_{C}(t_{f})$ implies
that $u_{C}(t)\geq0$ is a feasible solution of (\ref{cost2}) since
$u_{1}(t)<0$ for all $t\in\lbrack t_{0},t_{f}]$ cannot satisfy this condition.
Suppose that there exists some $[t_{1},t_{2})\subset\lbrack0,t_{f}]$ in which
the optimal solution satisfies $u_{C}^{\ast}(t)<0$. We will show that there
exists another control which would lead to a lower cost than $u_{C}^{\ast}%
(t)$. First, we construct a control $u_{C}^{1}(t)$ such that $u_{C}%
^{1}(t)=u_{C}^{\ast}(t)\geq0$ for $t\in\lbrack0,t_{1})\cup\lbrack t_{2}%
,t_{f}]$, $u_{C}^{1}(t)=0$ for $t\in\lbrack t_{1},t_{2})$. It is clear that
$u_{C}^{1}(t)$ will not violate the acceleration constraint
(\ref{vehicle_constraint}). However, the terminal position constraint is
violated. Therefore, we will construct $u_{C}^{2}(t)$, a variant of $u_{C}%
^{1}(t)$ as follows, and will show that $u_{C}^{2}(t)$ is feasible.

First, define
\begin{equation}
g_{C}(t)=x_{C}^{1}(t)+v_{C}^{1}(t)(t_{f}-t) \label{gc}%
\end{equation}
and note that $g_{C}(t)$ is continuous in $t$ since $x_{C}^{1}(t)$ and
$v_{C}^{1}(t)$ are continuous. Because $g_{C}(0)=\bar{x}_{C}(t_{f}%
)<x_{C}^{\ast}(t_{f})$ by assumption and $g_{C}(t_{f})=x_{C}^{1}(t_{f})\geq
x_{C}^{\ast}(t_{f})$, there exists some $t_{m}\in\lbrack0,t_{f}]$ such that
$g_{C}(t_{m})=x_{C}^{\ast}(t_{f})$. We can now construct $u_{C}^{2}(t)$ such
that $u_{C}^{2}(t)=u_{C}^{1}(t)\geq0$ for $t\in\lbrack0,t_{m})$, $u_{C}%
^{2}(t)=0$ for $t\in\lbrack t_{m},t_{f}]$. Observe that $x_{C}^{2}%
(t_{f})=g_{C}(t_{m})=x_{C}^{\ast}(t_{f})$. Moreover, based on its definition,
it is obvious that it will not violate the acceleration constraint. Next, we
show that the velocity constraint is also not violated. Suppose there exists
some time $t_{n}$ such that $v_{C}^{1}(t_{n})=v_{C\text{max}}$ {so that the
trajectory may include an arc over which $v_{C}^{1}(t)=$}$v_{C\text{max}}${.
}There are two cases:

\textbf{(a)} If $t_{n}\geq t_{m}$, we have $v_{C}^{1}(t_{m})\leq v_{C}%
^{1}(t_{n})$ because $u_{C}^{1}(t)\geq0$. Based on the definition of
$u_{C}^{1}(t)$, the maximal speed is $v_{C}^{1}(t_{m})$ and the velocity
constraint is not violated.

\textbf{(b)} If $t_{n}<t_{m}$, we have $v_{C}^{1}(t_{m})>v_{C}^{1}(t_{n})$.
Taking the time derivative of $g_{C}(t)$, we get $\dot{g}_{C}(t)=u_{C}%
^{1}(t)(t_{f}-t)\geq0$. Therefore,
\begin{equation}
g_{C}(t_{n})<g_{C}(t_{m})=x_{C}^{\ast}(t_{f}) \label{gcinequality}%
\end{equation}
We then construct a control $u_{C}^{3}(t)$ such that $u_{C}^{3}(t)=u_{C}%
^{\ast}(t)\geq0$ for $t\in\lbrack0,t_{1})\cup\lbrack t_{2},t_{n}]$, $u_{C}%
^{3}(t)=0$ for $t\in\lbrack t_{1},t_{2})\cup\lbrack t_{n},t_{f}]$. Note that
$t_{n}\geq t_{1}$ because $v_{C}^{\ast}(t)\neq v_{C\text{max}}$ when $t<t_{1}$
based on the definition of $u_{C}^{\ast}(t)$. If $t_{1}<t_{n}<t_{2}$, we
define $u_{C}^{3}(t)=0$ when $t>t_{1}$ as follows: $u_{C}^{3}(t)=u_{C}^{\ast
}(t)\geq0$, for $t\in\lbrack0,t_{1})$, $u_{C}^{3}(t)=0$, for $t\in\lbrack
t_{1},t_{f}]$. From the construction of $u_{C}^{3}(t)$, we have $x_{C}%
^{3}(t_{n})=x_{C}^{1}(t_{n})$, $v_{C}^{3}(t_{n})=v_{C}^{1}(t_{n})$ so that
\eqref{gcinequality} holds under $u_{1}^{3}(t)$, and $x_{C}^{3}(t_{f}%
)=g_{C}(t_{n})$. Because $v_{C}^{3}(t)=v_{C\text{max}}$ for $t\in\lbrack
t_{n},t_{f}]$ and $u_{C}^{3}(t)\geq u_{C}^{\ast}(t)$ for $t\in\lbrack0,t_{n}%
)$, it is clear that $x_{C}^{3}(t)\geq x_{C}^{\ast}(t)$ for all $t\in
\lbrack0,t_{f}]$. However, this contradicts \eqref{gcinequality} since
$x_{C}^{3}(t_{f})=g_{C}(t_{n})<x_{C}^{\ast}(t_{f})$. We conclude that
$t_{n}<t_{m}$ is not possible. In summary, we have proved that the speed
constraint will not be violated under the control $u_{C}^{2}(t)$.

Next, we show that $u_{C}^{2}(t)$ will also not violate the safety constraint.
Suppose that at time $t_{\sigma}\in\lbrack0,t_{f}]$, the safety constraint is
active under control $u_{C}^{2}(t)$, i.e., $x_{C}^{2}(t_{\sigma}%
)=x_{U}(t_{\sigma})-d_{C}$. Because $u_{C}^{2}(t)\geq0$, based on Lemma $2$,
it is straightforward to show that
\begin{equation}
\label{theorem4_safteyconstraint}x_{C}^{2}(t_{f})\geq x_{U}(t_{f})-d_{C}%
\end{equation}
Recall that, based on the definition of $u_{C}^{2}(t)$ and the condition that
$x_{C}(t_{f})<x_{U}(t_{f})-d_{C}$, we have $x_{C}^{2}(t_{f})=g_{C}%
(t_{m})=x_{C}^{\ast}(t_{f})<x_{U}(t_{f})-d_{C}$ which contradicts with
\eqref{theorem4_safteyconstraint}. Therefore, the safety constraint will never
be activated.

We conclude that $u_{C}^{2}(t)$ is a feasible solution. Moreover, under
$u_{C}^{2}(t)$ the cost is lower than that of $u_{C}^{\ast}(t)$ because
$u_{C}^{2}(t)$ contains a segment with $u_{C}^{2}(t)=0$ that contributes zero
cost in (\ref{cost2}) relative to $u_{C}^{\ast}(t)$. Therefore, the optimal
control $u_{C}^{\ast}(t)$ cannot contain any time interval with $u_{C}^{\ast
}(t)<0$.

{Finally, we use a similar argument as above to show that the safety
constraint will be inactive under the optimal control $u_{C}^{\ast}(t)$, that
is, $\eta_{5}^{\ast}(t)=0$.} Assume that at time $t_{\eta}\in(0,t_{f}]$,
$\eta_{5}^{\ast}(t_{\eta})>0$. Because the safety constraint is active at
$t_{\eta}$ and is not violated at $t_{f}$, vehicle $C$ must have
decelerated to relax the safety constraint. However, this violates the fact
that $u_{C}^{\ast}(t)\geq0$ as shown above. Therefore, we conclude that
$\eta_{5}^{\ast}(t)=0$ for al $t\in\lbrack0,t_{f}]$. This completes the
proof. $\blacksquare$

\textbf{Theorem $5$} [Case 2 in Fig. \ref{case_study}]: If $x_{C}(t_{f}%
)<\bar{x}_{C}(t_{f})<x_{U}(t_{f})-d_{C}$, then $u_{C}^{\ast}(t)\leq0$ and
$\eta_{5}^{\ast}(t)=0$.

\emph{Proof}: The proof is similar to that of Theorem $4$ and is
omitted. $\blacksquare$

\textbf{Theorem $6$} [Case 3 in Fig. \ref{case_study}] If ${x}_{C}(t_{f})<x_{U}(t_{f})-d_{C}<\bar{x}_{C}%
(t_{f})$, then $u_{C}^{\ast}(t)\leq0$.

\emph{Proof}: The proof is similar to Theorem $4$. The only difference is in
the way we prove that the constructed control $u_{C}^{2}(t)$ will not violate
the safety constraint. Suppose that there exists some $[t_{1},t_{2}%
)\subset\lbrack0,t_{f}]$ in which the optimal solution satisfies $u_{C}^{\ast
}(t)>0$. First, we construct a control $u_{C}^{1}(t)$ such that $u_{C}%
^{1}(t)=u_{C}^{\ast}(t)\leq0$ for $t\in\lbrack0,t_{1})\cup\lbrack t_{2}%
,t_{f}]$, $u_{C}^{1}(t)=0$ for $t\in\lbrack t_{1},t_{2})$. It is clear that
$x_{C}^{\ast}(t)\geq x_{C}^{1}(t)$, $v_{C}^{\ast}(t)\geq v_{C}^{1}(t)$,
$t\in\lbrack0,t_{f}]$. Considering the safety constraint in \eqref{cost2},
note that if $u_{C}^{\ast}(t)$ does not violate the safety constraint, then
neither does $u_{C}^{1}(t)$.

Using $g_{C}(t)$ defined in (\ref{gc}), note that $g_{C}(0)=x_{C}^{1}%
(0)+v_{C}^{1}(0)t_{f}=\bar{x}_{C}(t_{f})$ and $g_{C}(t_{f})=x_{C}^{1}(t_{f})$.
Since $x_{C}^{1}(t_{f})\leq x_{C}^{\ast}(t_{f})\leq\bar{x}_{C}(t_{f})$ and
$g_{C}(t)$ is continuous, there exists $t_{m}\in(0,t_{f})$ such that
$g_{C}(t_{m})=x_{C}^{\ast}(t_{f})$. Then, we construct $u_{C}^{2}%
(t)=u^{1}(t)\leq0$ for $t\in\lbrack0,t_{m})$ and $u_{C}^{2}(t)=0$ for
$[t_{m},t_{f}]$. Similar to the proof of Theorem $4$, $x_{C}^{2}(t_{f}%
)=g_{C}(t_{m})=x_{C}^{\ast}(t_{f})$. Since $u_{C}^{2}(t)=u_{C}^{1}(t)$ for
$t\in\lbrack0,t_{m})$, control $u_{C}^{2}(t)$ will not violate the safety
constraint when $t\leq t_{m}$. For $t>t_{m}$, we have $u_{C}^{2}(t)=0$ and
$x_{C}^{2}(t)$ is linear in $t$ with $v_{C}^{1}(t_{m})>0$. Moreover,
$x_{C}^{1}(t_{m})<x_{U}(t_{m})-d_{C}$ and $x_{C}^{2}(t_{f})={x}_{C}^{\ast
}(t_{f})<x_{U}(t_{f})-d_{C}$. We conclude that $u_{C}^{2}(t)$, $t\in\lbrack
t_{m},t_{f}]$, will not violate the safety constraint because the upper bound
of vehicle $C$'s safe position, $x_{U}(t)-d_{C}$, is also linear in $t$. Based
on the definition of $u_{C}^{2}(t)$, it is obvious that it will not violate
the acceleration constraint. We can then use the same argument as in the proof
of Theorem $4$ to show that $v_{C}(t_{m})\geq v_{C\min}$. Therefore,
$u_{C}^{2}(t)$ is a feasible solution. It is also obvious that the cost of
$u_{C}^{2}(t)$ is lower than that of $u_{C}^{\ast}(t)$ because $u_{C}^{2}(t)$
contains a segment with $u_{C}^{2}(t)=0$. Therefore, the optimal control
$u_{C}^{\ast}(t)$ cannot contain any time interval with $u_{C}^{\ast}(t)>0$.
This completes the proof. $\blacksquare$





Based on Theorems 4,5, \emph{Cases 1,2} in Fig. \ref{case_study} can be solved
without the safety constraint in (\ref{cost2}) since we have shown that
$\eta_{5}^{\ast}(t)=0$. Therefore, the optimal control is the same as that
derived for vehicles $1$ and $2$ in Theorems 2,3. This leaves only \emph{Case
3} to analyze. We proceed by first solving (\ref{cost2}) without the safety
constraint, so it reduces to the solution in Theorem 3, since we know that
$u_{C}^{\ast}(t)\leq0$. If a feasible optimal solution exists, then the
problem is solved. Otherwise, we need to re-solve the problem in order to
determine an optimal trajectory that includes at least one arc in which
$x_{U}(0)+v_{U}(0)t-x_{C}^{\ast}(t)-d_{C}=0$.

Based on Lemma $2$, there exists a time $\tau_{1}\in(0,t_{f})$ that satisfies
$v_{C}(\tau_{1})=v_{U}(0)$ and $x_{C}(\tau_{1})=x_{U}(0)+v_{U}(0)\tau
_{1}-d_{C}\equiv a$ (it is easy to see that there is at most one such
constrained arc, since $v_{C}(t)=v_{U}(0)$ as soon as this arc is entered.) We
then split problem \eqref{cost2} into two subproblems as follows:
\begin{equation}
\begin{aligned} \min_{u_C(t)}& \int_{0}^{\tau_1}\frac{1}{2}u_C^{2}(t)dt\\ \text{s.t. }& \eqref{vehicle_dynamics},\text{ }\eqref{vehicle_constraint},\text{ }x_C(\tau_1)=a,\text{ }v_C(\tau_1)=v_U(\tau_1),\text{ }t\in[0,\tau_1] \\ \end{aligned} \label{cost21}%
\end{equation}
\begin{equation}
\begin{aligned} &\min_{u(t)} \int_{\tau_1}^{t_f}\frac{1}{2}u_C^{2}(t)dt \text{ s.t. }\eqref{vehicle_dynamics},\text{ }\eqref{vehicle_constraint},\\ & \text{ }x_C(\tau_1)=a,\text{ }v_C(\tau_1)=v_U(\tau_1),\text{ }x_C(t_f)=x_{cf},\text{ }t\in[\tau_1,t_f]\\ \end{aligned} \label{cost22}%
\end{equation}
where (\ref{cost21}) has a fixed terminal time $\tau_{1}$ (to be determined),
position $a$, and speed $v_{U}(0)$, while (\ref{cost22}) has a fixed terminal
time $t_{f}$ and position $x_{C,f}$ with given $x_{C}(\tau_{1})=a$.

Let us first solve \eqref{cost21}. Since $u_{C}^{\ast}(t)\leq0$ and the
terminal speed is $v_{U}(0)$, only the acceleration constraint $u_{C\min
}-u_{C}\leq0$ can be active in $[0,\tau_{1}]$. Suppose that this constraint
becomes active at time $\tau_{2}<\tau_{1}$. Since $u_{C\min}-u_{C}$ is
independent of $t$, $x_{C}(t)$, and $v_{C}(t)$, it follows (see \cite{bryson2018applied}) that there are no discontinuities in the
Hamiltonian or the costates, i.e., $\lambda_{x}(\tau_{2}^{-})=\lambda_{x}%
(\tau_{2}^{+})$, $\lambda_{v}(\tau_{2}^{-})=\lambda_{v}(\tau_{2}^{+})$,
$H(\tau_{2}^{-})=H(\tau_{2}^{+})$. It follows from $H(\tau_{2}^{-})-H(\tau
_{2}^{+})=0$ and (\ref{Lagrangian_3}) that
\[
\lbrack u_{C}^{\ast}(\tau_{2}^{-})-u_{C}^{\ast}(\tau_{2}^{+})][\frac{1}%
{2}(u_{C}^{\ast}(\tau_{2}^{-}))+\frac{1}{2}(u_{C}^{\ast}(\tau_{2}%
^{+}))+\lambda_{v}(\tau_{2}^{-})]=0
\]
Therefore, either $u_{C}^{\ast}(\tau_{2}^{-})=u_{C}^{\ast}(\tau_{2}^{+})$ or
$u_{C}^{\ast}(\tau_{2}^{-})=-{\lambda_{v}(\tau_{2}^{-})}$ based on
\eqref{u_dynamics}. Either condition used in the above equation leads to the
conclusion that $u_{C}^{\ast}(\tau_{2}^{-})=u_{C}^{\ast}(\tau_{2}^{+})$, i.e.,
{$u_{C}^{\ast}(t)$ is continuous at $\tau_{2}$. }

Let us now evaluate the objective function in (\ref{cost21}) as a function of
$\tau_{1}$ and $a$, denoting it by $J_{1}(\tau_{1},a)$, under optimal control.
In view of \eqref{u_dynamics}, there are two cases.

\textbf{(a)} $u_{C}^{\ast}(t)=u_{C\text{min}}$ for $t\in\lbrack0,\tau_{2})$,
$u_{C}^{\ast}(t)=-{\lambda_{v}(t)}$ for $t\in\lbrack\tau_{2},\tau_{1}]$. As in
the proof of Theorem 2, the costate equations are $\dot{\lambda}%
_{v}(t)=-\lambda_{x}(t)$ and ${\dot{\lambda}}_{x}(t)=0$. Therefore,
$\lambda_{v}(t)=ct-b$ where $b,c$ are to be determined. It follows that%
\begin{equation}
u_{C}^{\ast}(t)=c(t-\tau_{2})+u_{C\text{min}}\text{, \ \ }t\in\lbrack\tau
_{2},\tau_{1}) \label{OptimalControl_1a}%
\end{equation}
and the following boundary conditions hold:%
\begin{align}
v_{C}(\tau_{2})  &  =v_{C}(0)+u_{C\text{min}}\tau_{2}\label{Subproblem 1a}\\
v_{C}(\tau_{1})  &  =v_{U}(0)=v_{C}(\tau_{2})+\int_{\tau_{2}}^{\tau_{1}%
}[c(t-\tau_{2})+u_{C\text{min}}]dt\nonumber\\
x_{C}(\tau_{2})  &  =x_{C}(0)+v_{C}(0)\tau_{2}+\frac{1}{2}u_{C\text{min}}%
\tau_{2}^{2}\nonumber\\
x_{C}(\tau_{1})  &  =a=x_{C}(\tau_{2})+\int_{\tau_{2}}^{\tau_{1}}[\frac{c}%
{2}t^{2}+(u_{C\text{min}}-c\tau_{2})(t-\tau_{2})\nonumber\\
&  -\frac{c}{2}\tau_{2}^{2}+v_{C}(0)+u_{C\text{min}}\tau_{2}]dt\nonumber
\end{align}
Using (\ref{OptimalControl_1a}) and (\ref{Subproblem 1a}) to eliminate $c$ and $\tau_2$ and then evaluate
$J_{1}(\tau_{1},a)$ in (\ref{cost21}) after some algebra yields: 
\begin{equation}
\begin{aligned} J_1(\tau_1,a)=&\frac{1}{2}u_\text{Cmin}(2v_U(0)-2v_C(0)-u_\text{Cmin}\tau_1)\\ &+\dfrac{2(v_U(0)-v_C(0)-u_\text{Cmin}\tau_1)^3}{9(a-x_C(0)-v_C(0)\tau_1-0.5u_\text{Cmin}\tau_1^2)}\\ \end{aligned} \label{u_B_Cost1}%
\end{equation}

\textbf{(b)} $u_{C}^{\ast}(t)=-{\lambda_{v}(t)}$ for $t\in\lbrack0,\tau_{2})$,
$u_{C}^{\ast}(t)=u_{C\text{min}}$ for $t\in\lbrack\tau_{2},\tau_{1}]$.
Proceeding as above, we get%
\begin{align}
v_{C}(\tau_{2})  &  =v_{C}(0)+\int_{0}^{\tau_{2}}[c(\tau_{2}-t)+u_{C\text{min}%
}]dt\label{Subproblem_1b}\\
v_{C}(\tau_{1})  &  =v_{U}(0)=v_{C}(\tau_{2})+(\tau_{1}-\tau_{2}%
)u_{C\text{min}}\nonumber\\
x_{C}(\tau_{2})  &  =x_{C}(0)+\int_{0}^{\tau_{2}}[v_{C}(0)+\frac{c}{2}%
t^{2}+u_{C\text{min}}t]dt\nonumber\\
x_{C}(\tau_{1})  &  =a=x_{C}(\tau_{2})+\int_{\tau_{2}}^{\tau_{1}}[v_{C}%
(\tau_{2})+u_{C\text{min}}(t-\tau_{2})]dt\nonumber
\end{align}
and, after some calculations, we obtain $J_{1}(\tau_{1},a)$ in (\ref{cost21}):
\begin{equation}
\begin{aligned} &J_1(\tau_1,a)=\frac{1}{2}u_\text{Cmin}(2v_U(0)-2v_C(0)-u_\text{Cmin}\tau_1)\\ &-\dfrac{2(v_U(0)-v_C(0)-u_\text{Cmin}\tau_1)^3}{9(a-x_C(0)-v_U\tau_1+0.5u_\text{Cmin}\tau_1^2)}\\ \end{aligned}
\end{equation}



Proceeding to the second subproblem (\ref{cost22}), note that the control at
the entry point of the constrained arc at time $\tau_{1}$ is no longer
guaranteed to be continuous. This problem is of the same form as the optimal
control problem for vehicle 2 in (\ref{OCP2_rev}) whose solution is given in
Theorem 3, except that initial conditions now apply at time $\tau_{1}$ as
given in (\ref{cost22}). Proceeding exactly as before, we can obtain the cost
$J_{2}(\tau_{1},a)$ under optimal control. Adding the two costs, we obtain
$J_{C}(\tau_{1},a)=J_{1}(\tau_{1},a)+J_{2}(\tau_{1},a)$\textbf{ }in
\eqref{cost2}. This results in a simple nonlinear programming problem whose
solution $(\tau_{1}^{\ast},a^{\ast})$ results from setting $\dfrac{\partial
J_{C}(\tau_{1},a)}{\partial\tau_{1}}=0$ and $\dfrac{\partial J_{C}(\tau
_{1},a)}{\partial a}=0$. Finally, the optimal control is the one corresponding
to $(\tau_{1}^{\ast},a^{\ast})$.


Based on our analysis, we find that \emph{Case 3}\ is the only one where the
safety constraint may become active. This provides an option to the vehicle
$C$ controller: if \emph{Case 3} applies, the maneuver may either be
implemented or it may be delayed until the conditions change to either one of
\emph{Cases 1,2} so as avoid the more complex situation that arises through
(\ref{cost21}),(\ref{cost22}).

\section{Simulation Results}

We provide simulation results illustrating the time and energy-optimal optimal
maneuver controller we have derived and compare its performance to a baseline
of human-driven vehicles. In what follows, we set the minimal and maximal
vehicle speeds to $1m/s$ and $33m/s$ respectively and the maximal acceleration
and deceleration to $3.3m/s^{2}$ and $-7m/s^{2}$ respectively. The
aggressiveness coefficients $\alpha_{i},i=1,2,C$ in \eqref{terminal time} are
all set to $\alpha_{i}=0.5$.

\begin{figure}[h]
\centering
\includegraphics[scale=0.22]{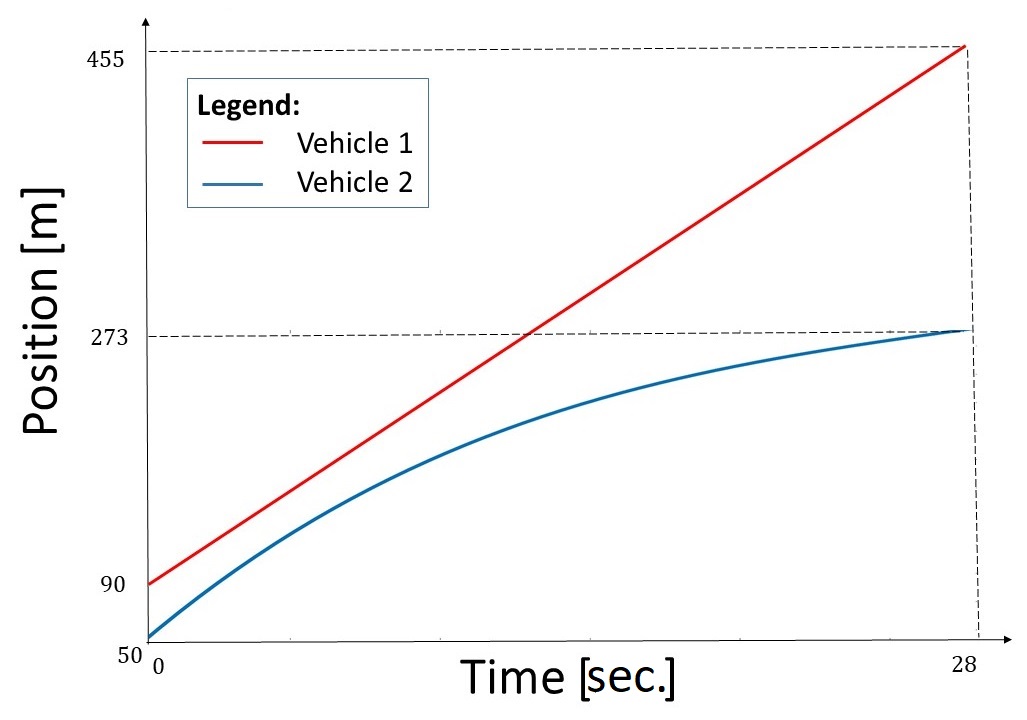} \caption{Optimal Position
trajectories of vehicle $1$ and $2$ in case ($1$) of Fig. \ref{case_study}.}%
\label{case1_12}%
\end{figure}

\begin{figure}[h]
\centering
\includegraphics[height=.11\textwidth,width=.23\textwidth]{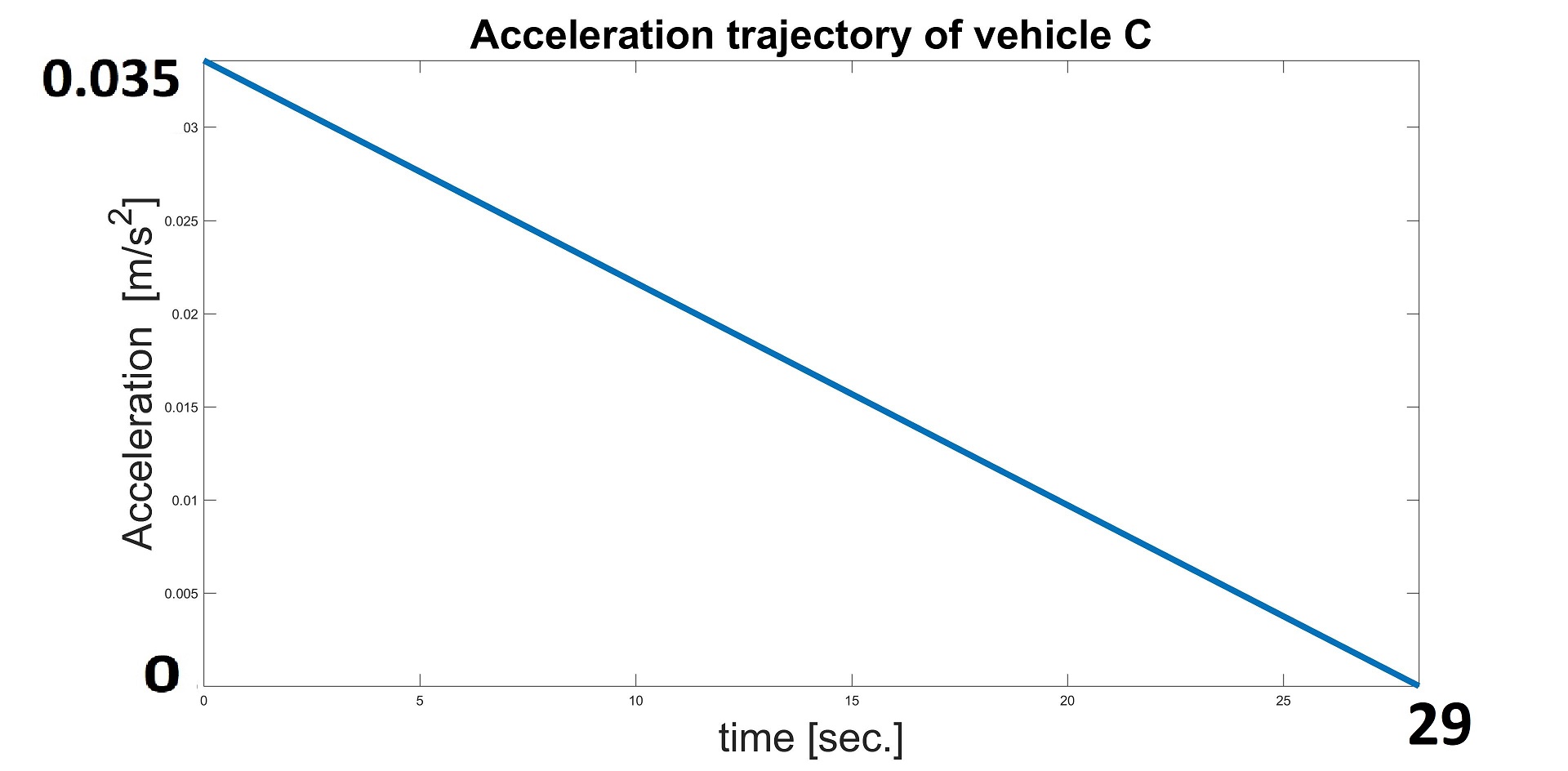}
\includegraphics[height=.11\textwidth,width=.23\textwidth]{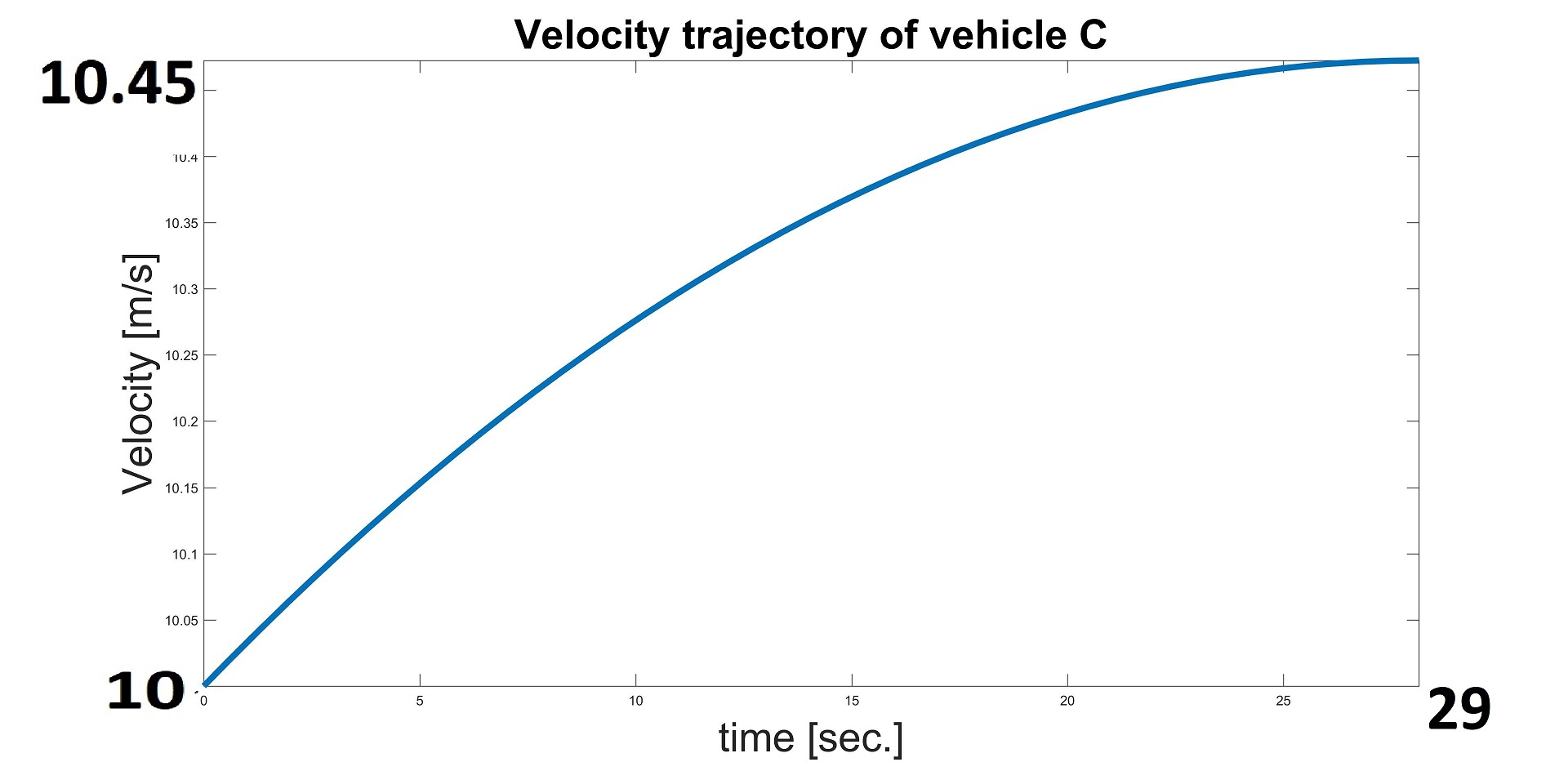}
\includegraphics[height=.17\textheight,width=.5\textwidth]{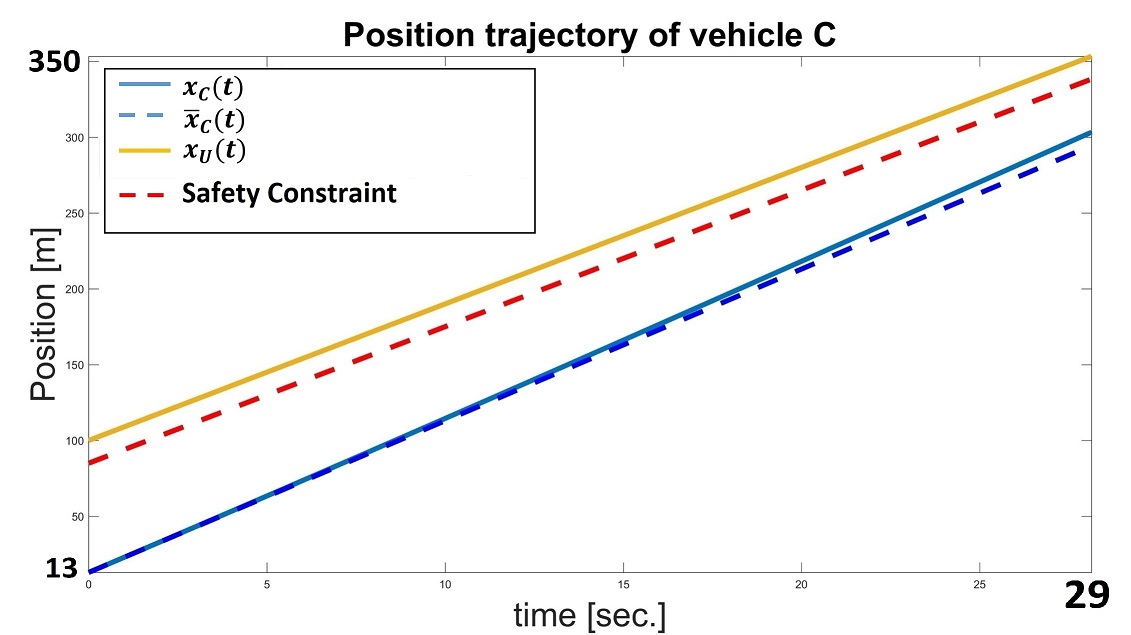} \caption{Optimal
trajectories of vehicle $C$ in case ($1$) of Fig. \ref{case_study}.}%
\label{case1_c}%
\end{figure}

\emph{Case 1}\textbf{ in Fig. \ref{case_study}. }We set $x_{1}(t_{0})=90m$,
$v_{1}(t_{0})=13m/s$, $x_{U}(t_{0})=100m$, $v_{U}(t_{0})=9m/s$, $x_{2}%
(t_{0})=50m$, $v_{2}(t_{0})=18m/s$ and $x_{C}(t_{0})=13m$, $v_{C}%
(t_{0})=10m/s$. Solving \eqref{terminal time}, we get $t_{f}=28.14s$ and after
solving \eqref{terminal position}, we obtain $x_{1}(t_{f})=455.8m$,
$x_{C}(t_{f})=303.24m$ and $x_{2}(t_{f})=273.24m$. Figs. \ref{case1_12}%
-\ref{case1_c} show the optimal trajectories of all controllable vehicles. In
Fig. \ref{case1_12}, vehicle $1$ is cruising with a constant velocity which
contributes a zero value to the cost in \eqref{OCP1}, while the velocity of
vehicle $2$ decreases to create space for vehicle $C$ to change lanes. The
optimal trajectory of vehicle $C$ in Fig. \ref{case1_c} is obtained without
considering the safety constraint because of Theorem $4$. Vehicle $C$ keeps on
accelerating and the safety distance constraint is never
violated.

\begin{figure}[h]
\centering
\includegraphics[scale=0.22]{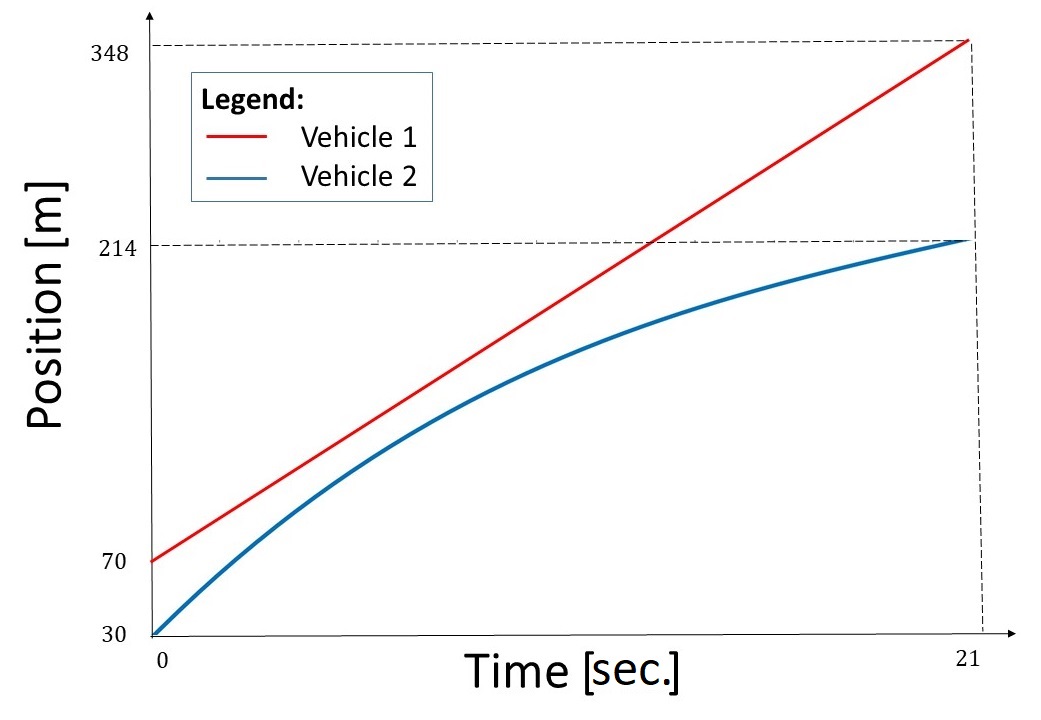} \caption{Optimal Position
trajectories of vehicle $1$ and $2$ in case ($2$) of Fig. \ref{case_study}.}%
\label{case2_12}%
\end{figure}

\begin{figure}[h]
\centering
\includegraphics[height=.11\textwidth,width=.23\textwidth]{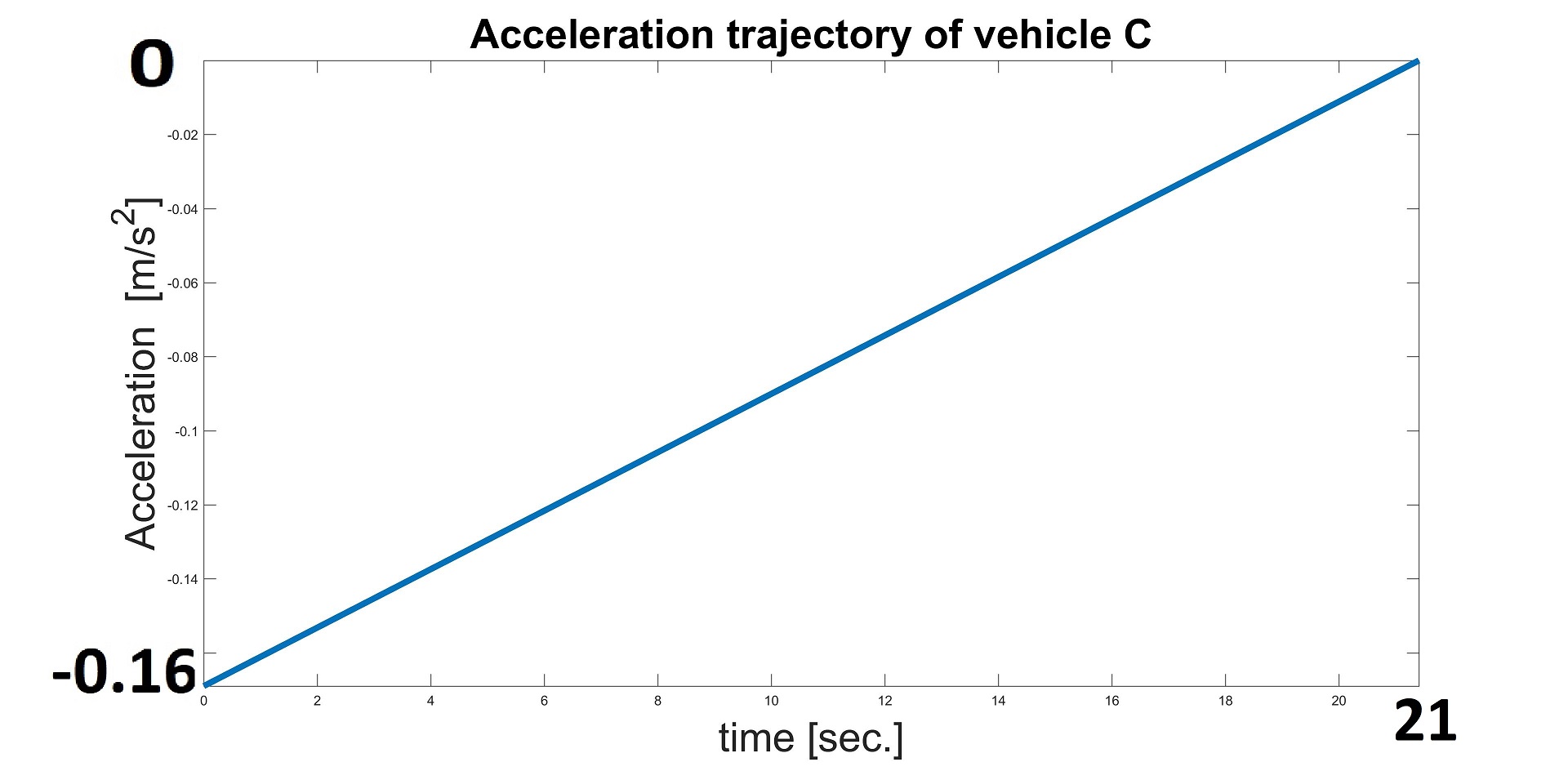}
\includegraphics[height=.11\textwidth,width=.23\textwidth]{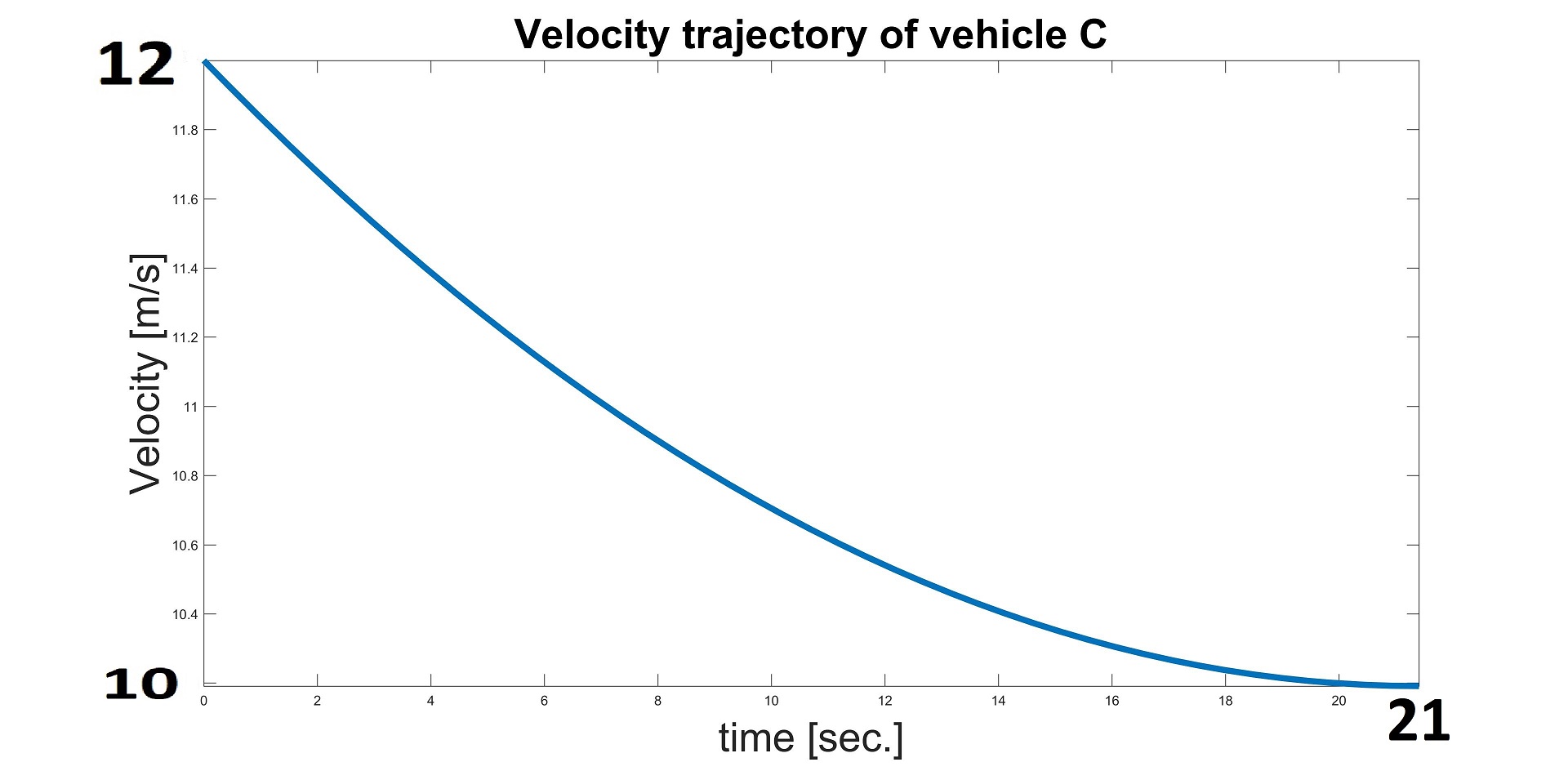}
\includegraphics[height=.17\textheight,width=.5\textwidth]{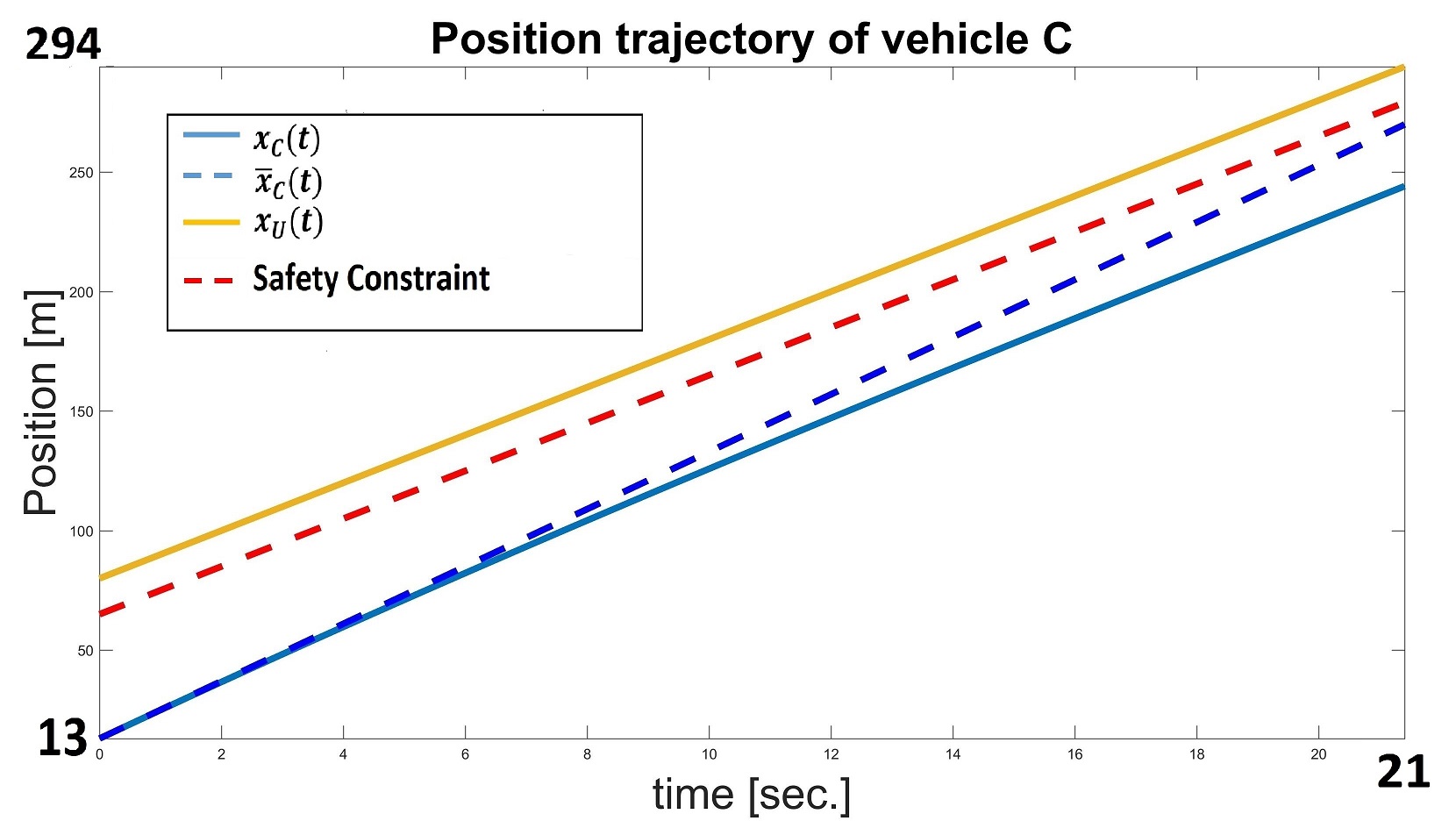} \caption{Optimal
trajectories of vehicle $C$ in case ($2$) of Fig. \ref{case_study}.}%
\label{case2_c}%
\end{figure}

\emph{Case 2}\textbf{ in Fig. \ref{case_study}. }We set $x_{1}(0)=70m$,
$v_{1}(0)=13m/s$, $x_{2}(0)=30m$, $v_{2}(0)=18m/s$, $x_{C}(0)=13m$,
$v_{C}(0)=12m/s$, $x_{U}(0)=80m$, $v_{U}(0)=10m/s$. Solving
\eqref{terminal time} and \eqref{terminal position}, we get $t_{f}=21.4s$ and
$x_{1}(t_{f})=348.37m$, $x_{2}(t_{f})=214.13m$, $x_{C}(t_{f})=244.13m$. Figure
\ref{case2_12} shows the optimal trajectories of vehicles $1$,$2$ in which $1$
is cruising with a constant speed and the associated energy cost is zero,
while the velocity of vehicle $2$ decreases. Figure \ref{case2_c} shows the
optimal trajectory of vehicle $C$ which, once again, is obtained without
considering the safety constraint based on Theorem $5$. Vehicle $C$
decelerates to ensure it satisfies its terminal position while the safety
constraint is never violated.

\begin{figure}[h]
\centering
\includegraphics[scale=0.22]{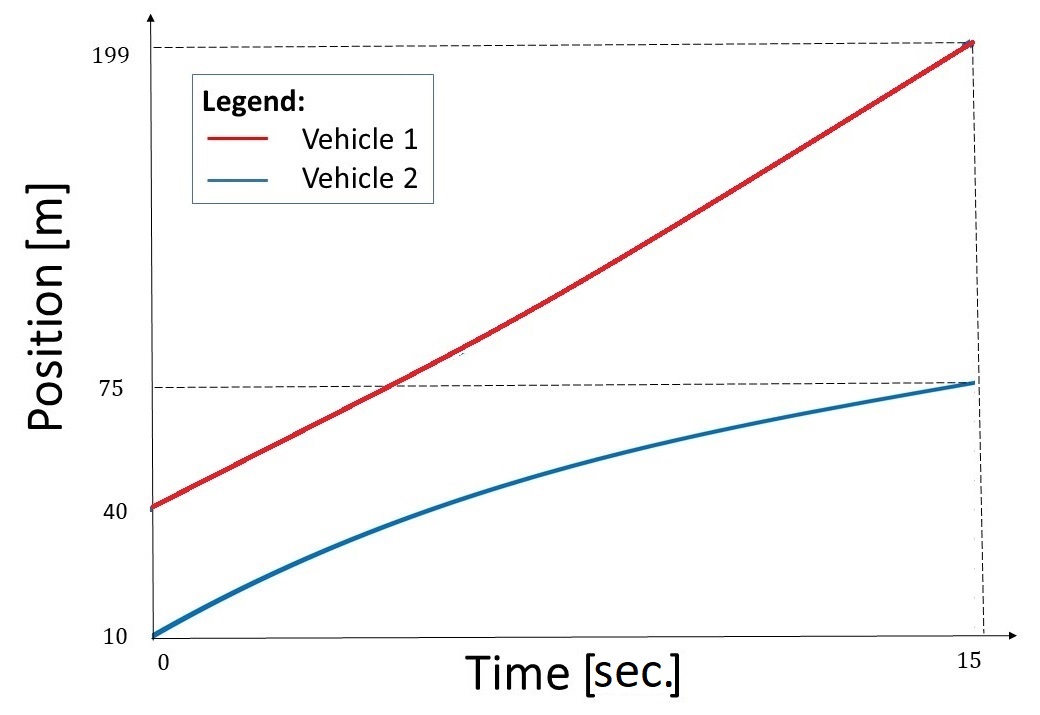} \caption{Optimal Position
trajectories of vehicle $1$ and $2$ in case ($3$) of Fig. \ref{case_study}.}%
\label{case3_12}%
\end{figure}

\begin{figure}[h]
\centering
\includegraphics[height=.11\textwidth,width=.23\textwidth]{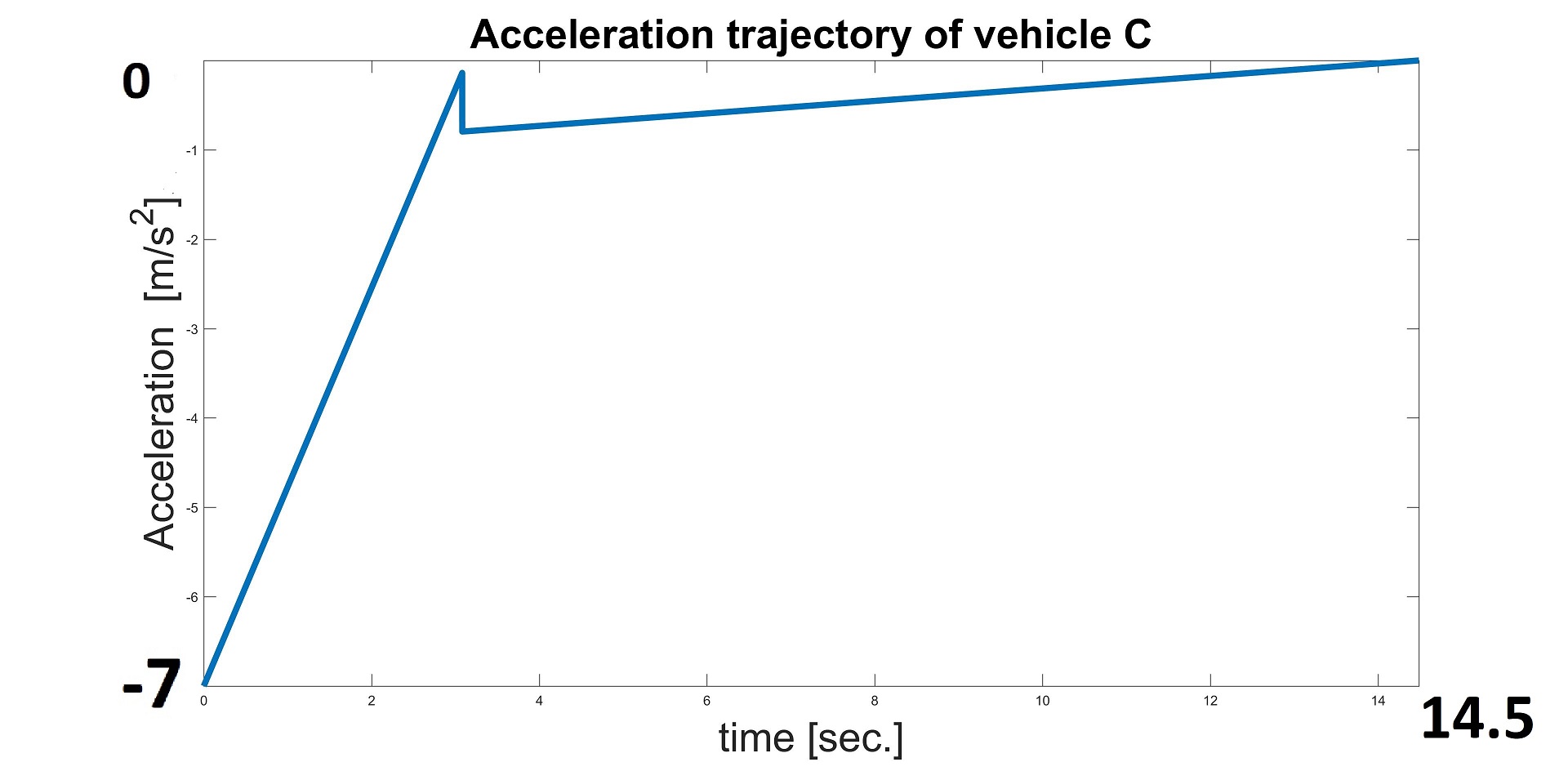}
\includegraphics[height=.11\textwidth,width=.23\textwidth]{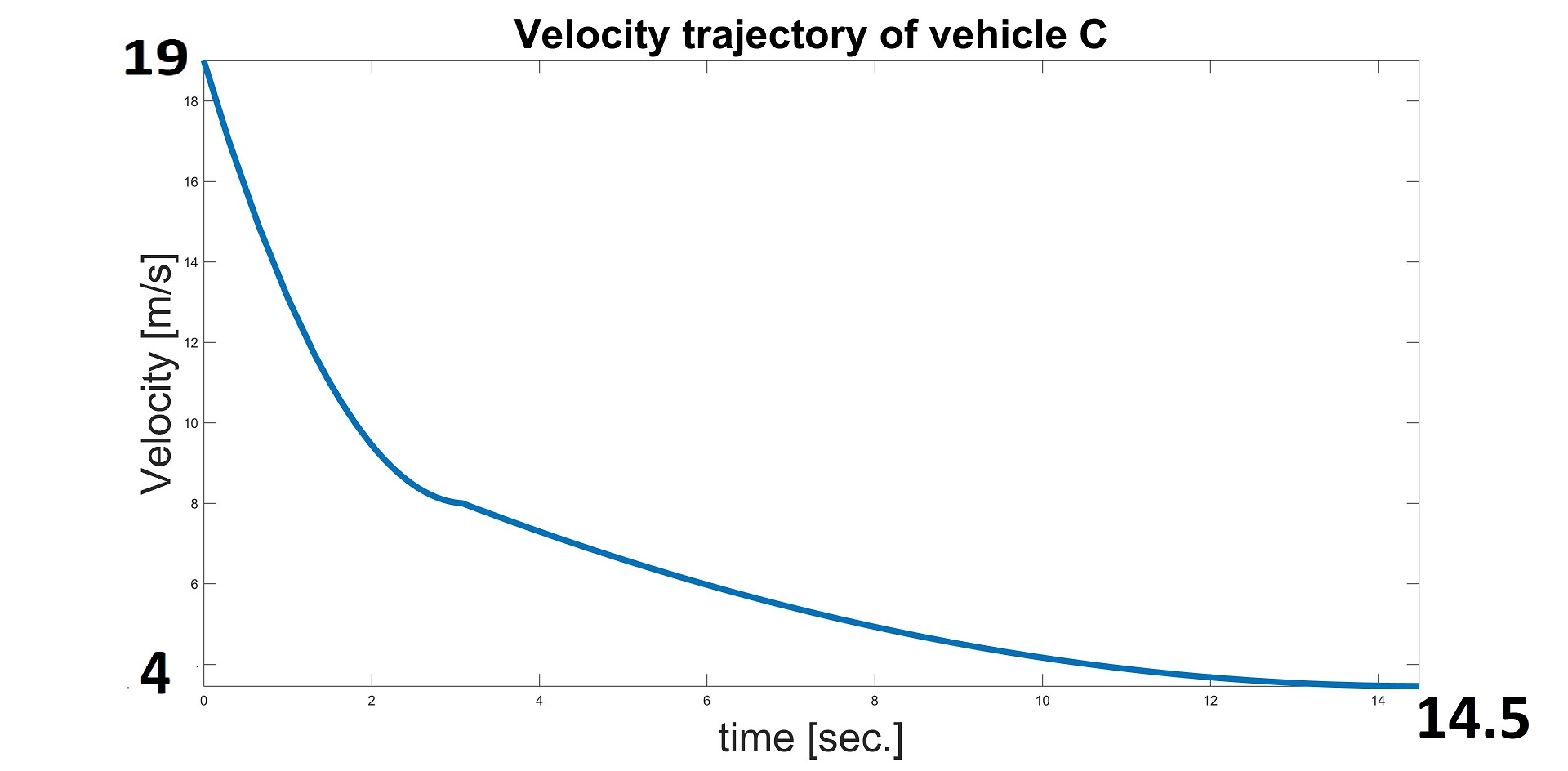}
\includegraphics[height=.17\textheight,width=.5\textwidth]{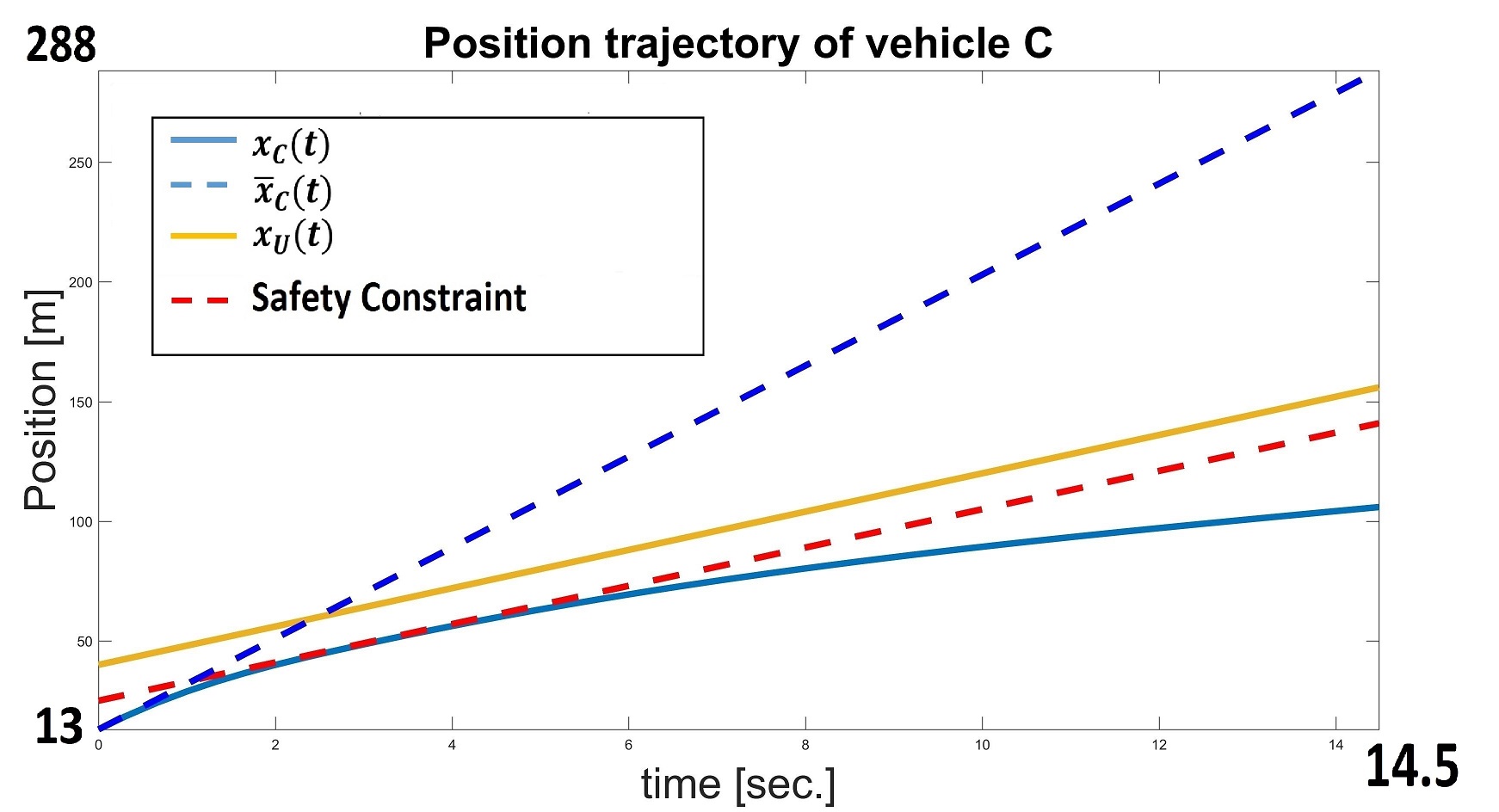} \caption{Optimal
trajectories of vehicle $C$ in case ($3$) of Fig. \ref{case_study}.}%
\label{case3_c}%
\end{figure}

\emph{Case 3}\textbf{ in Fig. \ref{case_study}. }We set $x_{1}(0)=40m$,
$v_{1}(0)=11m/s$, $x_{U}(0)=40m$, $v_{U}(0)=8m/s$ $x_{2}(0)=10m$,
$v_{2}(0)=23m/s$, $x_{C}(0)=13m$, $v_{C}(0)=19m/s$. Solving
\eqref{terminal time} and \eqref{terminal position}, we get $t_{f}=14.49s$ and
$x_{1}(t_{f})=199.37m$, $x_{2}(t_{f})=75m$, $x_{C}(t_{f})=105.9m$. The optimal
trajectories of vehicles $1$,$2$ are shown in Fig. \ref{case3_12}. In this
case, vehicle $1$ accelerates and vehicle $2$ decelerates in order to create
space for vehicle $C$. For vehicle $C$, we first solve the optimal control
problem (\ref{cost2}) without considering the safety constraint and find that
it actually becomes active. Therefore, we proceed with the two subproblems
(\ref{cost21}) and (\ref{cost22}) to derive the true optimal trajectories. We
obtained $a^{\ast}=43m$ and $\tau_{1}^{\ast}=3.2s$, and Fig. \ref{case3_c}
shows the optimal trajectory of vehicle $C$. Observe that $C$ decelerates over
the maneuver and the safety distance constraint is active at $\tau_{1}^{\ast
}=3.2s$ when there is a jump in the acceleration trajectory. Following that,
vehicle $C$ continues decelerating until it reaches its terminal position.

\textbf{Comparison of optimal maneuver control and human-driven vehicles}. We
use standard car-following models in the commercial SUMO simulator to simulate
a lane change maneuver implemented by human-driven vehicles with the
requirement that vehicle $C$ changes lanes between vehicles $1$ and $2$. We
considered all cases in Fig. \ref{case_study} with both CAVs and human-driven
vehicles sharing the same initial states as shown in Table \ref{tab:1}.
\begin{table}[ptb]
\caption{Initial states of vehicles}%
\label{tab:1}%
\vspace*{-\baselineskip}
\par
\begin{center}
\resizebox{.48\textwidth}{.030\textheight}{
\begin{tabular}
[c]{c||c|c|c|c|c|c|c|c|c}\hline
\backslashbox{Cases}{States} & {$x_{1}(0)[m]$} & {$v_{1}(0)[m/s]$} &
$x_{2}(0)[m]$ & $v_{2}(0)[m/s]$ & $x_{C}(0)[m]$ & $v_{C}(0)[m/s]$ &
$x_{U}(0)[m]$ & $v_{U}(0)[m/s]$ & $d_{C}[m]$\\\hline\hline
(1) & 95 & 13 & 0 & 18 & 13 & 10 & 120 & 9 & 30\\\hline
(2) & 120 & 13 & 30 & 18 & 13 & 16 & 100 & 10 & 30\\\hline
(3) & 100 & 11 & 10 & 23 & 213 & 19 & 290 & 8 & 30\\\hline\hline
\end{tabular}}
\end{center}
\end{table}The associated energy consumption is shown in Table \ref{tab:2} and
provides evidence of savings in the range $43-59\%$ over all three cases.
\begin{table}[ptb]
\caption{
	ENERGY COMPARISON: CAVs vs HUMAN-DRIVEN VEHICLES}%
\label{tab:2}%
\vspace*{-\baselineskip}
\par
\begin{center}
\resizebox{.48\textwidth}{.030\textheight}{
\begin{tabular}
[c]{c||c|c|c}\hline
\backslashbox{Cases}{Energy Consumption} & CAVs & {Human-driven Vehicles} &
Improvement \\\hline\hline
(1) & 6.8 & 16.4 & 59\%  \\\hline
(2) & 23.0 & 46.0 & 50\%  \\\hline
(3) & 59.5 & 103.5 & 43\%  \\\hline\hline
\end{tabular}}
\end{center}
\end{table}

\section{Conclusion and Future Work}

We used an optimal control framework to derive time and energy-optimal
policies for a CAV cooperating with neighboring CAVs to implement a highway
lane change maneuver. We optimize the maneuver time and subsequently minimize
the associated energy consumption of all cooperating vehicles in this
maneuver. Our solution is limited to the first step of the complete maneuver,
i.e., all three cooperating vehicles adjust their positions before the
lane-changing vehicle makes the lane shift. Our ongoing work aims to complete
this step. In addition, we plan to incorporate a \textquotedblleft
comfort\textquotedblright\ factor in the problem by minimizing any resulting
jerk and adopt a more general velocity-varying safety distance constraint.

\bibliographystyle{ieeetran}
\bibliography{cmp}

\begin{thebibliography}{10}
\providecommand{\url}[1]{#1}
\csname url@samestyle\endcsname
\providecommand{\newblock}{\relax}
\providecommand{\bibinfo}[2]{#2}
\providecommand{\BIBentrySTDinterwordspacing}{\spaceskip=0pt\relax}
\providecommand{\BIBentryALTinterwordstretchfactor}{4}
\providecommand{\BIBentryALTinterwordspacing}{\spaceskip=\fontdimen2\font plus
\BIBentryALTinterwordstretchfactor\fontdimen3\font minus
  \fontdimen4\font\relax}
\providecommand{\BIBforeignlanguage}[2]{{%
\expandafter\ifx\csname l@#1\endcsname\relax
\typeout{** WARNING: IEEEtran.bst: No hyphenation pattern has been}%
\typeout{** loaded for the language `#1'. Using the pattern for}%
\typeout{** the default language instead.}%
\else
\language=\csname l@#1\endcsname
\fi
#2}}
\providecommand{\BIBdecl}{\relax}
\BIBdecl

\bibitem{varaiya1993smart}
P.~Varaiya, ``Smart cars on smart roads: problems of control,'' \emph{IEEE
  Trans. on Automatic Control}, vol.~38, no.~2, pp. 195--207, 1993.

\bibitem{zhao2018accelerated}
D.~Zhao, X.~Huang, H.~Peng, H.~Lam, and D.~J. LeBlanc, ``Accelerated evaluation
  of automated vehicles in car-following maneuvers,'' \emph{IEEE Trans. on
  Intelligent Transportation Systems}, vol.~19, no.~3, pp. 733--744, 2018.

\bibitem{wang2016cooperative}
M.~Wang, W.~Daamen, S.~P. Hoogendoorn, and B.~van Arem, ``Cooperative
  car-following control: Distributed algorithm and impact on moving jam
  features,'' \emph{IEEE Trans. on Intelligent Transportation Systems},
  vol.~17, no.~5, pp. 1459--1471, 2016.

\bibitem{wang2015game}
M.~Wang, S.~P. Hoogendoorn, W.~Daamen, B.~van Arem, and R.~Happee, ``Game
  theoretic approach for predictive lane-changing and car-following control,''
  \emph{Transportation Research Part C: Emerging Technologies}, vol.~58, pp.
  73--92, 2015.

\bibitem{nilsson2015longitudinal}
J.~Nilsson, M.~Br{\"a}nnstr{\"o}m, E.~Coelingh, and J.~Fredriksson,
  ``Longitudinal and lateral control for automated lane change maneuvers,''
  \emph{Proc. of 2015 American Control Conf.}, pp. 1399--1404, 2015.

\bibitem{bax2014road}
C.~Bax, P.~Leroy, and M.~P. Hagenzieker, ``Road safety knowledge and policy: A
  historical institutional analysis of the {Netherlands},''
  \emph{Transportation Research part F: Traffic Psychology and Behaviour},
  vol.~25, pp. 127--136, 2014.

\bibitem{you2015trajectory}
F.~You, R.~Zhang, G.~Lie, H.~Wang, H.~Wen, and J.~Xu, ``Trajectory planning and
  tracking control for autonomous lane change maneuver based on the cooperative
  vehicle infrastructure system,'' \emph{Expert Systems with Applications},
  vol.~42, no.~14, pp. 5932--5946, 2015.

\bibitem{werling2010optimal}
M.~Werling, J.~Ziegler, S.~Kammel, and S.~Thrun, ``Optimal trajectory
  generation for dynamic street scenarios in a frenet frame,'' \emph{Proc. of
  2010 IEEE Intl. Conf. on Robotics and Automation}, pp. 987--993, 2010.

\bibitem{bevly2016lane}
D.~Bevly, X.~Cao, M.~Gordon, G.~Ozbilgin, D.~Kari, B.~Nelson, J.~Woodruff,
  M.~Barth, C.~Murray, A.~Kurt \emph{et~al.}, ``Lane change and merge maneuvers
  for connected and automated vehicles: A survey,'' \emph{IEEE Trans. on
  Intelligent Vehicles}, vol.~1, no.~1, pp. 105--120, 2016.

\bibitem{nilsson2017lane}
J.~Nilsson, M.~Br{\"a}nnstr{\"o}m, E.~Coelingh, and J.~Fredriksson, ``Lane
  change maneuvers for automated vehicles,'' \emph{IEEE Trans. on Intelligent
  Transportation Systems}, vol.~18, no.~5, pp. 1087--1096, 2017.

\bibitem{luo2016dynamic}
Y.~Luo, Y.~Xiang, K.~Cao, and K.~Li, ``A dynamic automated lane change maneuver
  based on vehicle-to-vehicle communication,'' \emph{Transportation Research
  Part C: Emerging Technologies}, vol.~62, pp. 87--102, 2016.

\bibitem{mahjoub2017learning}
H.~N. Mahjoub, A.~Tahmasbi-Sarvestani, H.~Kazemi, and Y.~P. Fallah, ``A
  learning-based framework for two-dimensional vehicle maneuver prediction over
  v2v networks,'' \emph{Proc. of 15th IEEE Intl. Conf. on Dependable, Autonomic
  and Secure Computing}, pp. 156--163, 2017.

\bibitem{kazemi2018learning}
H.~Kazemi, H.~N. Mahjoub, A.~Tahmasbi-Sarvestani, and Y.~P. Fallah, ``A
  learning-based stochastic {MPC} design for cooperative adaptive cruise
  control to handle interfering vehicles,'' \emph{IEEE Trans. on Intelligent
  Vehicles}, vol.~3, no.~3, pp. 266--275, 2018.

\bibitem{kamal2013model}
M.~A.~S. Kamal, M.~Mukai, J.~Murata, and T.~Kawabe, ``Model predictive control
  of vehicles on urban roads for improved fuel economy,'' \emph{IEEE Trans. on
  Control Systems Technology}, vol.~21, no.~3, pp. 831--841, 2013.

\bibitem{katriniok2013optimal}
A.~Katriniok, J.~P. Maschuw, F.~Christen, L.~Eckstein, and D.~Abel, ``Optimal
  vehicle dynamics control for combined longitudinal and lateral autonomous
  vehicle guidance,'' \emph{Proc. of 2013 Control Conf.}, pp. 974--979, 2013.

\bibitem{lam2013cooperative}
S.~Lam and J.~Katupitiya, ``Cooperative autonomous platoon maneuvers on
  highways,'' \emph{Proc. of 2013 IEEE/ASME Intl. Conf. on Advanced Intelligent
  Mechatronics}, pp. 1152--1157, 2013.

\bibitem{li2017optimal}
B.~Li, Y.~Zhang, Y.~Ge, Z.~Shao, and P.~Li, ``Optimal control-based online
  motion planning for cooperative lane changes of connected and automated
  vehicles,'' \emph{Proc. of 2017 IEEE/RSJ Intl. Conf. on Intelligent Robots
  and Systems}, pp. 3689--3694, 2017.

\bibitem{vogel2003comparison}
K.~Vogel, ``A comparison of headway and time to collision as safety
  indicators,'' \emph{Accident Analysis \& Prevention}, vol.~35, no.~3, pp.
  427--433, 2003.

\bibitem{meng2018optimal}
X.~Meng and C.~G. Cassandras, ``Optimal control of autonomous vehicles for
  non-stop signalized intersection crossing,'' \emph{Proc. of 57th IEEE Conf.
  on Decision and Control}, pp. 6988--6993, 2018.

\bibitem{bryson2018applied}
A.~E. Bryson, \emph{Applied optimal control: optimization, estimation and
  control}.\hskip 1em plus 0.5em minus 0.4em\relax Routledge, 2018.

\end{thebibliography}

\end{document}